# Chapter 18. The Elusive Origin of Mercury

Denton S. Ebel and Sarah T. Stewart

**Denton S. Ebel**
Department of Earth and Planetary Sciences, American Museum of Natural History, New York
Department of Earth and Environmental Sciences, Columbia University, New York, NY
Graduate School and University Center of City University of New York
(debel@amnh.org)

**Sarah T. Stewart**
Department of Earth and Planetary Sciences, University of California, Davis, CA
(sts@ucdavis.edu )

**Abstract**
    The MESSENGER mission sought to discover what physical processes determined Mercury's high metal to silicate ratio. Instead, the mission has discovered multiple anomalous characteristics about our innermost planet. The lack of FeO and the reduced oxidation state of Mercury's crust and mantle are more extreme than nearly all other known materials in the solar system. In contrast, moderately volatile elements are present in abundances comparable to the other terrestrial planets. No single process during Mercury's formation is able to explain all of these observations. Here, we review the current ideas for the origin of Mercury's unique features. Gaps in understanding the innermost regions of the solar nebula limit testing different hypotheses. Even so, all proposed models are incomplete and need further development in order to unravel Mercury's remaining secrets.

## 18.1 Introduction

The MESSENGER mission to Mercury was driven by several scientific goals. A principal question was "What planetary formational processes led to Mercury's high ratio of metal to silicate" (Solomon et al., 2001, 2007; Chapter 1). This ratio is the prime anomaly for Mercury relative to other planetary bodies, but by no means the only one. In this chapter, we review Mercury's anomalous chemistry in the context of astrophysical disk processes and planet formation as currently understood. We examine previous and current hypotheses for Mercury's origin in light of the new data from MESSENGER. Finally, we discuss the potential for Mercury-like planets in the rapidly emerging catalog of extrasolar planetary systems. A discussion of the evolution of Mercury as a planet, focusing on processes occurring after many or all of its anomalous properties were established, is given in Chapter 19.

## 18.2 How anomalous is Mercury?

By far the smallest terrestrial planet, Mercury is about half the mass of Mars. Mercury is also locked in a 3:2 spin–orbit resonance with the Sun. These factors may have affected Mercury's collisional environment (Chapter 9). Here, we focus on data returned by the MESSENGER mission that may provide insights into Mercury's formation. A surprising result from the MESSENGER mission is that Mercury's large ratio of metal to silicate is not accompanied by extreme depletions in volatile elements compared with the other terrestrial planets. The planet's mantle is much more chemically reduced, and less S-depleted, than Earth's mantle, and also compared with carbonaceous chondrite meteorites, the usual Solar System standard. These MESSENGER findings have stimulated many experimental studies bearing on the possible S and Si abundances in Mercury's large core. We begin by summarizing the evidence returned by the MESSENGER mission in the context of chemical anomalies.

### 18.2.1 Density

It has long been known that Mercury's zero-pressure, or uncompressed, density is significantly greater than that of Earth – 5000 versus 4400 kg/m$^3$ according to Urey (1950), and 5300 versus 4100 kg/m$^3$ according to Mahoney (2014) – and greater than the grain density of most meteorite parent bodies (Figure 18.1). It can be assumed that Mercury's density is attributable to a high abundance of iron and heavy, siderophile (iron-loving) elements, relative to silicon, magnesium, and other lithophile (rock-loving) elements. Upon planetary differentiation, the siderophile elements accompany iron that sinks to form the planetary core, and the lithophile elements form the mantle (Righter, 2003). Given that Mercury differentiated in the same manner as other planets, it is, therefore, the core/mantle mass ratio that is anomalous. The metal/silicate or Fe/Si ratios of primitive meteorites may be considered reference points for processes that led to Mercury's high density (Chapter 2).

In 1988, Chapman noted that the origin of Mercury was "the least well-understood topic" in the compendium of Mercury knowledge (Chapman, 1988). He divided solutions to Mercury's metal-rich composition into two categories. Either the composition reflects primordial radial gradients of "orderly" chemical fractionation and dynamic processes, or it results from "chaotic" evolutionary processes that catastrophically overprinted primordial signatures (Wetherill, 1994). Indeed, the relative roles of such processes have implications for the origin of all the planets in this and other planetary systems.

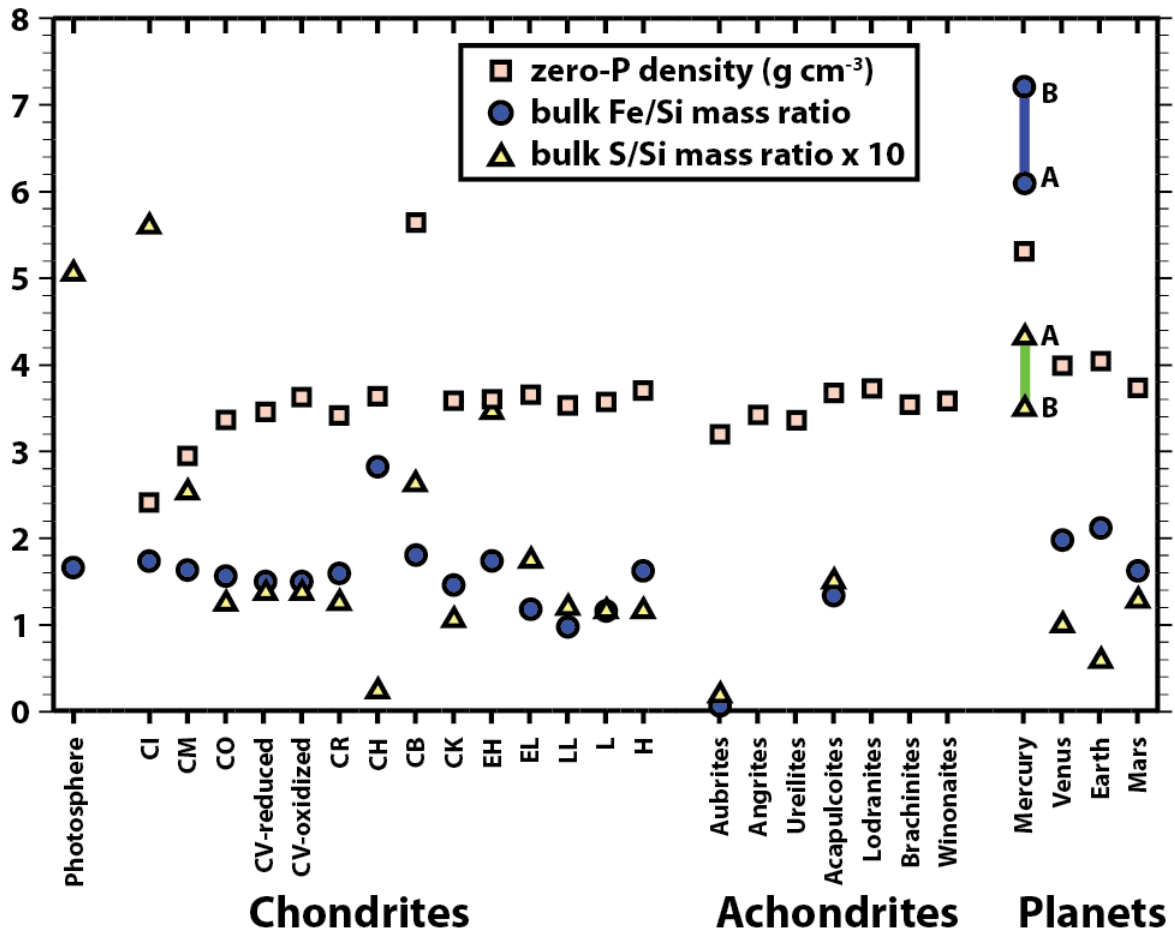

**Figure 18.1**. Mean grain densities of chondrites and achondrites and zero-pressure densities of planets (squares), bulk Fe/Si mass ratios (circles), and bulk S/Si mass ratios (multiplied by 10, triangles). Grain density is the bulk material density in the absence of porosity. Mercury values are calculated as described in the text, showing ranges (vertical bars) for Fe/Si and S/Si for cases **A** and **B**, as marked. Density data are from Macke (2012, meteorites) and Lodders and Fegley (1998, planets). Fe/Si data are from Wasson and Kallemeyn (1988): CM, CO, CV, EH, EL, LL, L, H; Jarosewich (1990): CB; Lodders (2003): solar photosphere; Lodders et al. (2009): CI; Lodders and Fegley (1998): CR, CH, CK, Earth, aubrites, acapulcoites, Venus, Mars.

The CI chondrite meteorites contain most non-volatile elements in ratios (relative to Si or Mg) nearly identical to those measured by spectroscopy in the solar photosphere, which is thought to represent the bulk composition of the Solar System (Lodders, 2003; Lodders et al., 2009; Sneden et al., 2009). Approximately chondritic ratios of the major rock-forming elements (Si, Mg, Fe, S, Ca, Al, and Ni) characterize Venus, Earth, and Mars (Righter et al., 2006). A rough calculation shows that if Mercury were once a differentiated planet with near-chondritic bulk chemistry it must have lost a substantial mass of silicate mantle to yield its present density. The thickness of a shell of such additional silicate would be ~1480 km to yield a present Mercury radius of 2440 km with 74% of its mass in a metallic, Si-free core. An initially chondritic Mercury would have been about 530 km larger in radius than Mars (3389.5 km mean radius) prior to stripping ~64% of its mass, all from its mantle. Alternatively, Mercury was never chondritic in major elements and owes its anomalous density to as yet unknown nebular fractionation processes.

### 18.2.2 FeO-free silicates

The second major chemical anomaly for Mercury is the planet's extremely reduced nature. Here, we compare Mercury to the oxidation/reduction, or redox, state of other Solar System bodies and materials. Earth-based astronomical measurements of Mercury's spectral reflectance have long indicated a very low FeO content in surface silicates (Vilas, 1988; Sprague and Roush, 1998; Robinson and Taylor, 2001; Warell and Blewett, 2004). Microwave emissivity measurements (Mitchell and de Pater, 1994) precluded the possibility of widespread reduction to iron nanophases by space weathering (Domingue et al., 2014). The MESSENGER mission has confirmed the extreme depletion of Fe in any form on Mercury's surface and also in material excavated from depth (Murchie et al., 2015; Chapter 8). Indeed, the common crustal minerals olivine, pyroxene, and feldspar lack sufficient $Fe^{2+}$ to yield spectral features, indicating that Fe is present in reduced form as metal or sulfides (Klima et al., 2013; Izenberg et al., 2014; Chapter 8). Calcium in most terrestrial and extraterrestrial rocks forms oxides, but a positive correlation between Ca and S in the measurements by MESSENGER's X-Ray Spectrometer (XRS) suggest the occurrence of CaS (oldhamite) on Mercury's surface (Weider et al., 2012).

The chemical availability of oxygen, known as oxygen fugacity $f(O_2)$, is commonly described on a logarithmic scale relative to the $f(O_2)$ of a standard equilibrium reaction buffer, e.g., between iron metal and wüstite (~FeO), which is denoted by IW (Figure 18.2). As $f(O_2)$ decreases in an otherwise closed chemical system, more Si can dissolve into molten or solid Fe-rich metal, and more S can dissolve in silicate melts. From a variety of criteria based primarily on experimental metal-silicate partitioning and the MESSENGER XRS measurements of surface sulfur and iron abundances (Nittler et al., 2011), McCubbin et al. (2012) determined that the oxygen fugacity of Mercury's interior lies between a low of IW-6.3 and a high of IW-2.6, that is, $10^{-2.6}$ to $10^{-6.3}$ below IW. The high end of this range is an extremely generous upper limit (cf., Zolotov et al., 2013). The results from MESSENGER also stimulated other experiments that led to similar findings. Namur et al. (2016) parameterized S and metal solubility in magma compositions representative of mercurian lavas, and given mantle/core equilibrium they calculated $f(O_2)$ = IW-5.4±0.4 ($10^{-5.0}$ to $10^{-5.8}$ below IW; Figure 8.2). Earth, on the other hand, was not as reduced during its accretion (e.g., Frost et al., 2008), and modern mid-ocean ridge basalts record upper mantle $f(O_2)$ at $10^2$ above IW (IW+2), near the quartz-fayalite-magnetite (QFM) buffer (Cottrell and Kelley, 2011; Fig. 8.2). The mantle source of venusian lavas is inferred to have $fO_2$ similar to the Earth's upper mantle (Wadhwa, 2008).

The known or inferred $f(O_2)$ values for a broad range of Solar System materials are summarized in Figure 18.2. Mercury is the most reduced planet and more reduced than all measured early Solar System materials except, possibly, the enstatite chondrite and enstatite achondrite meteorites (aubrites) and a small class of Ca-, Al-rich inclusions (CAIs) in some chondritic meteorites (Figure 18.2; Beckett, 1986). Recent geochemical thermodynamic modeling suggests that as Earth and Mars accreted, their oxidation states progressively increased (Righter et al., 2008; Wood et al., 2009; Rubie et al., 2015; Badro et al., 2015; Righter et al., 2016). Earth's lower mantle, Mars, and Vesta (represented by diogenite meteorites) all record mantle $f(O_2)$ near the iron-wüstite buffer curve (Figure 18.2) (Ghosal et al., 1998; Wadhwa, 2001; Frost et al., 2008; Szymanski et al., 2010; Tuff et al., 2013). The FeO content of silicates and the Si content of metals are proxies for the $f(O_2)$ during their formation. The silicates in pallasite meteorites record $f(O_2)$ similar to that in Earth's lower mantle (Figure 18.2), with olivine, $(Mg,Fe)_2SiO_4$, ranging from 10 to 20 mol% $Fe_2SiO_4$ (11–19 wt% FeO) (Righter et al., 1990) and very low Si in metal. The most Si-rich iron meteorite, Horse Creek, contains 2.5 wt% Si in metal (Buchwald, 1975). Thus the meteorites that represent cores or lower mantles of early differentiated planetesimals all record higher $f(O_2)$ than Mercury.

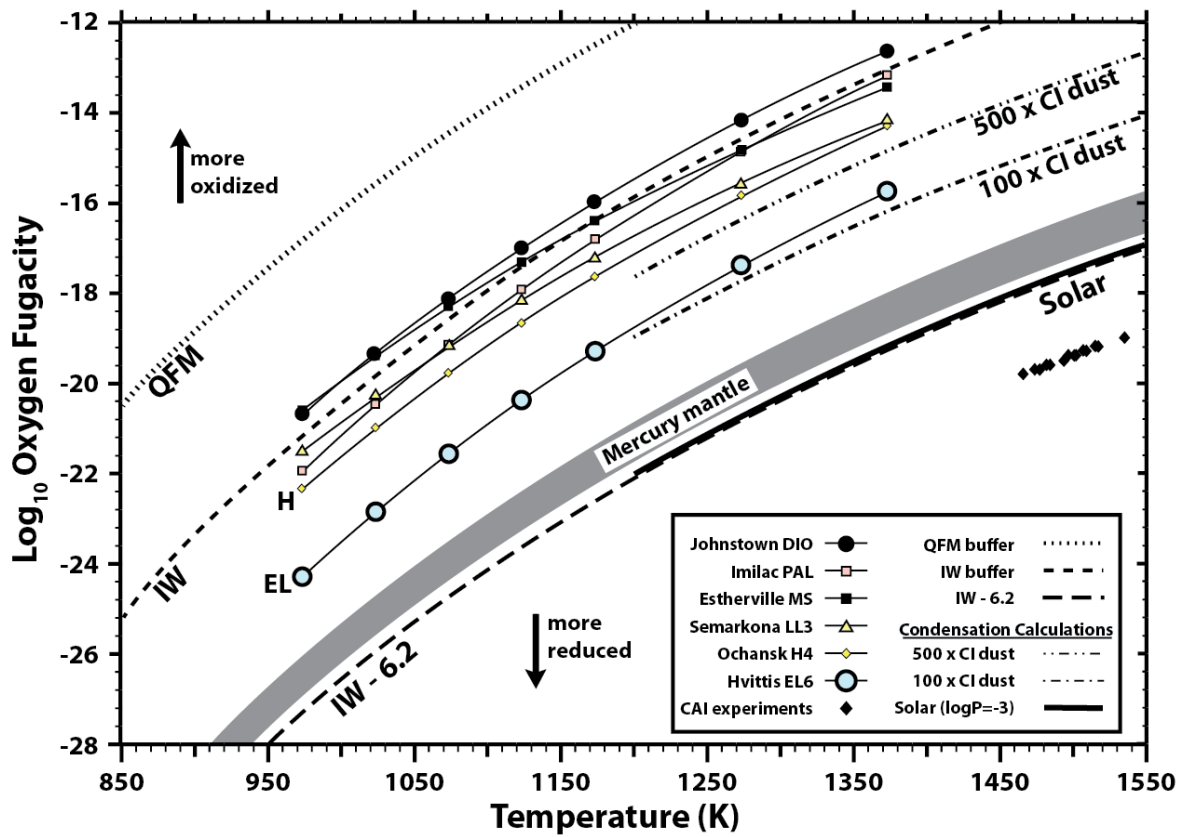

**Figure 18.2**. Intrinsic oxygen fugacity ($fO_2$) for selected meteorites (Brett and Sato, 1984: LL3, H4, EL6 chondrites; Hewins and Ulmer, 1984: mesosiderites MS, diogenites DIO; Righter et al., 1990: pallasites PAL) and during condensation of possible early disk vapors (solar composition and solar composition enriched in CI dust by factors of 500 and 1000), all at 100 Pa ($10^{-3}$ bar) total pressure (Ebel and Grossman, 2000). Buffer curves are for quartz-fayalite-magnetite (QFM), iron-wüstite (Fe-FeO, IW), and IW-6.2, which coincides with the calculated solar condensation curve. Condensation curves are nearly independent of the total pressure of the calculation (Ebel and Grossman, 2000). Results from the low-$fO_2$ experiments of Beckett (1986) are denoted by solid diamonds. The band shows the range (IW-5.4±0.4) of $fO_2$ estimated by Namur et al. (2016) for mantle source regions of >75% of mercurian lavas, on the basis of experiments.

The intrinsic $f(O_2)$, inferred from thermodynamic analysis of measured mineral compositions, has been calculated for some chondrite classes. The intrinsic $f(O_2)$ measured for the ordinary chondrites least equilibrated on their parent bodies (LL3, H4) range as low as IW-1; however, the equilibrated enstatite chondrite Hvittis (EL6) reaches IW-3 (Figure 18.2; Brett and Sato, 1984). The reduced, metal-rich CH and CB chondrites contain primarily <5 wt% FeO in olivine (Weisberg et al., 2001), and the CH chondrites contain metal grains, a very few of which reach 8 wt% Si (Weisberg et al., 1988). The enstatite achondrites (aubrites) are similar to the enstatite chondrites in their degree of reduction; for example, Mount Egerton contains 2.1 wt% Si in kamacite metal (Wasson and Wai, 1970). A new subgroup of metal-rich chondrites has recently been described, with ~22 vol% metal and nearly FeO-free silicates (Weisberg et al., 2015). The equilibrated enstatite chondrites contain ubiquitous Si-bearing metal (~3.2 wt% in EH4-5, ~1.6 wt% in EL6), no olivine, and <0.9 wt% FeO in enstatite (Keil, 1968; Weisberg and Kimura, 2012, their Fig. 5). The higher Si content of metal in equilibrated EH chondrites indicates that they are more chemically reduced than the EL6 chondrites. Of all the rocky bodies in the Solar System, only some enstatite chondrites and Mercury approach the low $f(O_2)$ of a vapor of solar composition that condenses solids along a buffer curve of IW-6.2 (Ebel and Grossman, 2000).

Vigarano-like carbonaceous (CV) chondrites contain once-melted (igneous) CAIs that crystallized Al-, Ti-rich calcic pyroxene (fassaite) containing $Ti^{3+}$ (Simon et al., 2007). Experiments by Beckett (1986) constrained the stability conditions of such pyroxenes in CAI-like melts to $\log_{10} f(O_2)$ ~ IW-8 at temperature $T$ in the range 1470 K < $T$ < 1540 K (Figure 18.2). The substantial $Ti^{3+}$ in these pyroxenes indicates highly reduced crystallization conditions in their parental melts, which also record the oldest radiometric ages of all Solar System materials (Russell et al., 2006). These CAIs are thought to have formed near the Sun, in highly reducing environments, and then been dynamically transported outward to accrete with more oxidized chondritic material (e.g., Krot et al., 2009; Brownlee, 2014). The host CV chondrites for these rare pyroxene-rich igneous CAIs are younger and not especially reduced, and it is only the CAIs in these meteorites that record such low $f(O_2)$. These CAIs offer tantalizing evidence that local regions in the earliest nebula were highly reduced.

In summary, there exist reservoirs of undifferentiated early Solar System material that record highly reduced conditions during formation and accretion of solids in the early nebula. A preponderance of evidence indicates that these materials formed by gradual, orderly processes. The vast majority of sampled materials, however, record much more oxidizing conditions of formation, accretion, and differentiation than does Mercury.

### 18.2.3 Elemental Abundances

The primary constraints on Mercury's formation come from composition information (Lewis, 1988; Boynton et al., 2007; Chapter 2) and inferred internal structure (Chapter 4). Before MESSENGER, our best estimate of Mercury's bulk composition was largely based on a theory of planet formation in which volatility varied with radial distance from the early Sun (Morgan and Anders, 1980; Lodders and Fegley, 1998). MESSENGER has changed the equations markedly. While the MESSENGER mission has established tighter bounds on Mercury's reduced mantle, it has also shown that Mercury is not anomalous in several important ways. The less volatile lithophile elements Si, Ca, Al, and Mg are all broadly chondritic in the crust (Weider et al., 2015; Chapter 2). Crustal values must be extrapolated to a bulk planetary composition, as is done to establish model bulk compositions for the other terrestrial planets.

The elemental ratios in Figures 18.1 and 18.3 are derived from the mean mantle source compositions inferred for the source regions of magmas that solidified to form the northern smooth plains (NSP) and intercrater plains and heavily cratered terrain (IcP-HCT) and given in Chapter 2 (Table 2.2). Two cases are plotted. Case **A** is consistent with Chapter 2, adding 7 wt% S and 2 wt% C (Section 2.6; Namur et al., 2016) to a 340-km-thick mantle shell of the mean NSP plus IcP-HCT "mass balance" composition (Table 2.2), with an FeS lower mantle shell of 80 km thickness and a 2020-km-radius core with 5% Ni, 1.5% S, 0.5% C, 8 wt% Si, and the remainder Fe (Table 2.3; Hauck et al., 2013; Chapter 4). Case **B** replaces all but 0.01 km of the FeS shell with the same mantle as **A** and sequesters 3 wt% S and only 4 wt% Si in the core. Case **B** brings the elemental Si/Mg and S/Mg ratios to near EH chondrite levels and also decreases Fe/Mg. This calculation may be within the uncertainty regarding (1) the existence of an FeS shell, (2) Mercury's internal thermal profile, (3) the olivine/pyroxene ratio of the mantle source, assumed near the low end of the range for lherzolites in Chapter 2. The calculation reveals another significant compositional anomaly, the high planetary Si/Mg mass ratio caused even by a conservative inference of only 4 wt% Si in the core. However, current estimates of the light elements in the core are based on single-stage core formation models under the assumption of thermodynamic equilibrium between core and mantle. More complete core formation models should reexamine the implications of a high bulk Si/Mg ratio for Mercury.

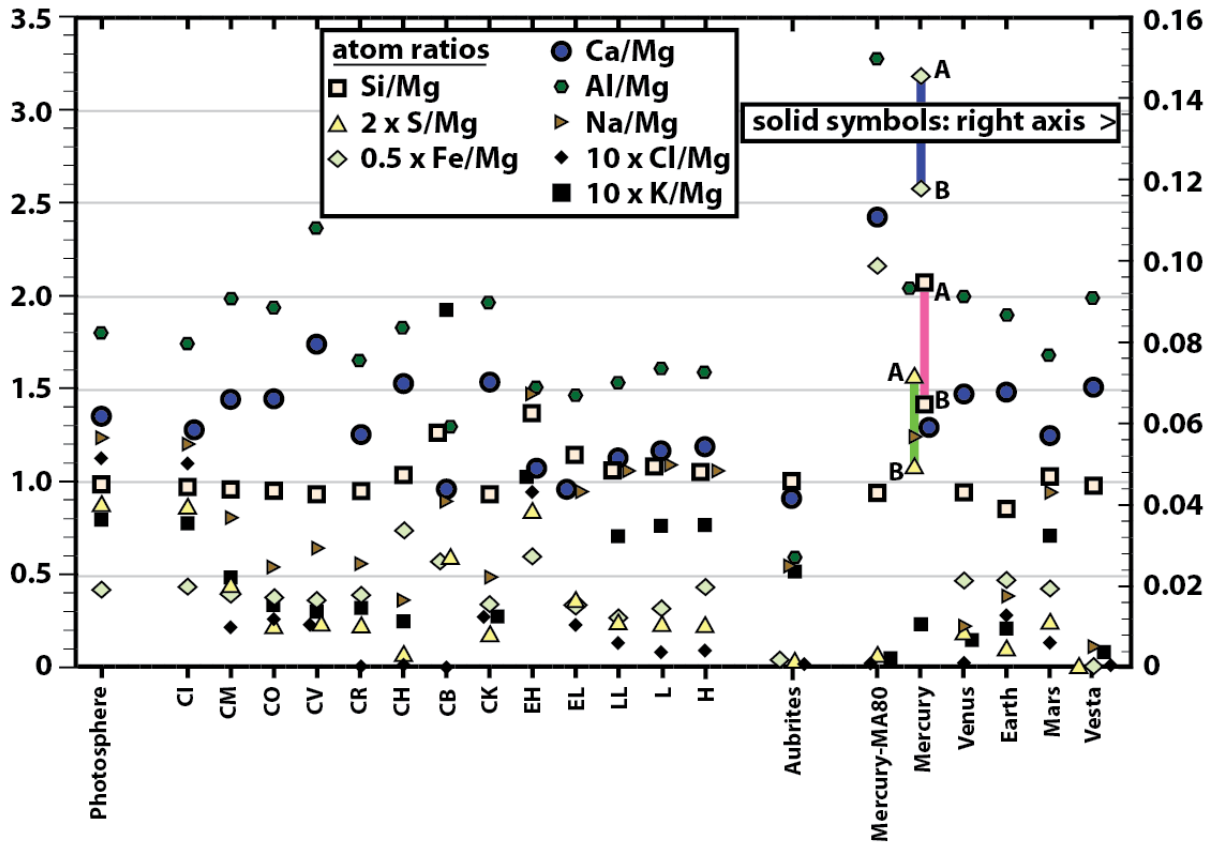

**Figure 18.3**. Mass ratios of lithophile elements in the solar photosphere, meteorites, and selected planets. Mercury values are calculated as described in the text, showing ranges (vertical bars) for Fe/Mg, S/Mg, and Si/Mg for cases **A** and **B**, as marked. "Mercury-MA80" and Venus values are from Morgan and Anders (1980). Earth and Mars are from Lodders and Fegley (1998; after Kargel and Lewis, 1993; and Lodders and Fegley, 1997; respectively). Other data sources are as in Figure 18.1. Si/Mg, 2×S/Mg, and 0.5×Fe/Mg (open symbols) refer to the left-hand axis. No Cl/Mg ratios for bulk CB, aubrites, or Mercury were calculated.

Earth, most primitive meteorites, and probably Venus, if not Mars, are all depleted in sulfur and other volatile elements relative to the chondritic (CI) reference composition. The average surface S/Si mass ratio of Mercury (0.092±0.015, Evans et al., 2012) is comparable to or higher than bulk Earth estimates. The volatile element depletion of the bulk Earth (dashed line in Figure 18.4) is known from the lithophile element abundances in mantle-derived rocks. Earth's bulk, planetary S/Mg atomic ratio is inferred to be ~7% of chondritic ("1" in Figure 18.4), with most S thought to be in the core, which is why S plots so far below the dashed line. By contrast, experimental and thermodynamic constraints (Namur et al., 2016) indicate that Mercury's bulk S/Mg mass ratio (Figure 18.3) is nearly identical to that of the solar photosphere, the most primitive chondrites (CI), and also EH chondrites.

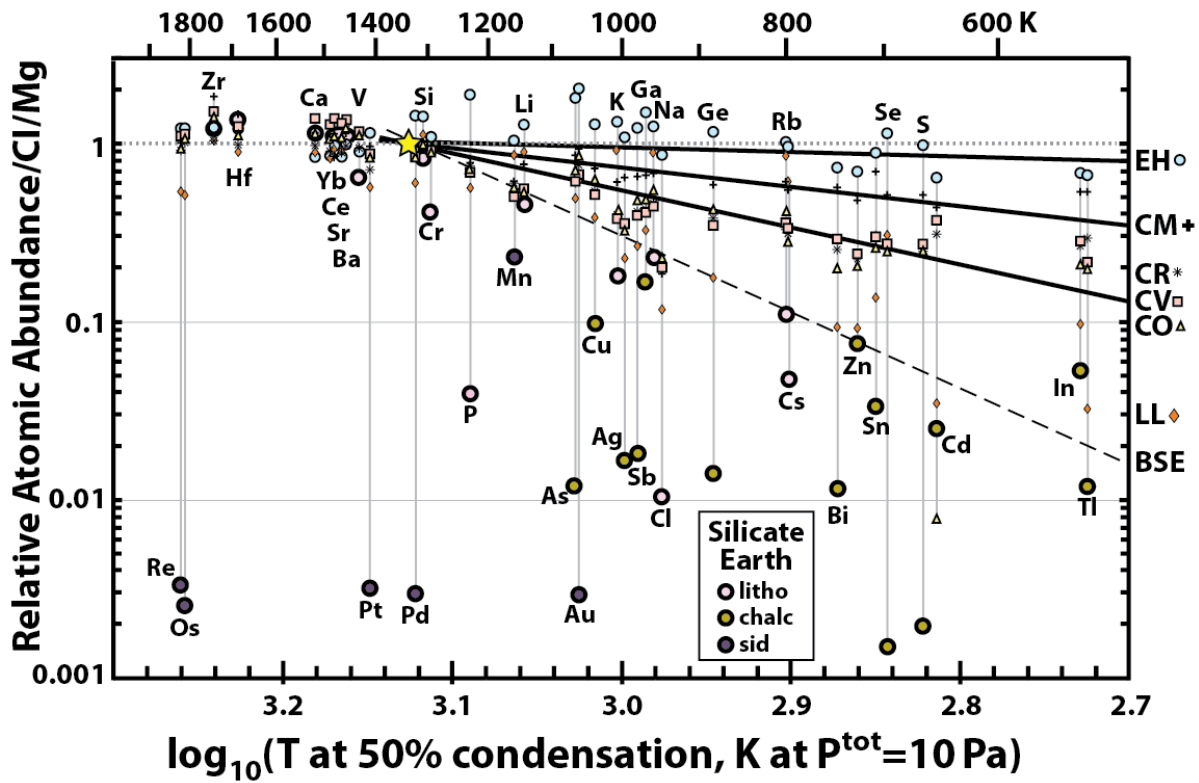

**Figure 18.4**. Volatile depletion of Solar System reservoirs. Bulk silicate Earth (BSE, large circles), LL (diamonds), CO (triangles), CV (squares), CR (asterisks), CM (crosses), and EH (small circles) chondrite atomic abundances of selected elements, normalized to Mg (star) and CI chondrite (dotted line at a relative atomic abundance of 1) are plotted against the temperature $T_{50\%}$ at which they are 50% condensed from a vapor of solar composition at 10 Pa ($10^{-4}$ bar) total pressure (Lodders, 2003). Earth data are from McDonough (2014); CI chondrite from Lodders et al. (2009); CM, CO, CV, EH, LL from Wasson and Kallemeyn (1988); and CR from Lodders and Fegley (1998). Trend lines for bulk silicate Earth (dashed), CV, CM, and EH are estimated. In accord with Goldschmidt's (1937) geochemical classification, bulk silicate Earth's lithophile, chalcophile, and siderophile elements are distinguished by shading.

Mercury's crustal K/Th and K/U ratios are slightly higher than those of Earth and Mars (Peplowski et al., 2011, 2012; Figure 2.3). The ratios of the moderately volatile ($T_{50\%}$ ~ 1000 K, see Figure 18.4) element K to the highly refractory ($T_{50\%}$ > 1600 K) elements Th or U reflect differences in the abundances of moderately volatile elements between planetary bodies. These elements are all highly incompatible in crystals that remain behind when partial melts rise to planetary surfaces, so their ratios are preserved in volcanic rocks that result from mantle melting processes. However, if some U and Th were incorporated into the core, which is possible under highly reduced conditions, the total mercurian inventory of K may be lower than that inferred from surface measurements of K/U and K/Th, under the assumption of chondritic mantle K, U, and Th (Malavergne et al., 2010; McCubbin et al., 2012). Furthermore, the crustal Cl/K ratio is close to chondritic (Evans et al., 2015; Chapter 2), as is the crustal Na/Si ratio (Evans et al., 2012; Peplowski et al., 2014). Overall, Cl and Na appear to be present in near-chondritic abundances. Mercury's abundances of moderately volatile K, Na, and Cl relative to refractory Mg are inferred to be most similar to those measured for the EH enstatite chondrites (Figure 18.4).

In summary, Mercury is anomalously enriched in Fe, S, and possibly also Si, relative to the other terrestrial planets (Figure 18.3). Its apparent bulk Na and possibly Cl compositions are also enriched above chondritic, whereas K is depleted, but less so than for Earth (Figure 18.3). A model for Mercury's origin must elevate Si-rich Fe metal compared with Mg-silicates and retain volatile S and Na at near-chondritic values relative to Mg, while maintaining abundance ratios of K, Ca, and Al that are similar to those for the other terrestrial planets. A reasonable inference from these observations is that the formation processes that led to the depletion of moderately volatile elements in planets compared with chondrites were decoupled from the origin of the large metal fraction of Mercury.

### 18.2.4 Surface reflectance

The surface reflectance of airless bodies depends upon chemical composition and regolith maturity. Mercury has a lower global reflectance than the Moon, but matures at a rate more rapid by as much as a factor of 4 (Robinson et al., 2008; Braden and Robinson, 2013; Chapter 8). Comparison of immature surfaces indicates that the difference in reflectance is likely due primarily to differences in the compositions of their surfaces. Darkening agents on the Moon include Fe- and Ti-bearing phases, but Mercury's surface is depleted in both elements. Low-reflectance material (LRM) on Mercury is up to 30% lower in reflectance than the global mean, but LRM is not enriched in either Fe or Ti, and Fe concentration does not correlate with reflectance (Weider et al., 2012, 2014; Murchie et al., 2015). LRM may represent excavated mid to lower crustal material (Ernst et al., 2010), so the darkening agent may represent a major component of the silicate portion of Mercury.

Gamma-Ray Spectrometer (GRS) measurements are consistent with 0 to 4.1 wt% C on Mercury's surface (Peplowski et al., 2015, 2016; Chapter 2). High thermal neutron count rates measured by the MESSENGER Neutron Spectrometer (NS) correlate with LRM, consistent with LRM C abundances 1–3 wt% greater than in surrounding higher-reflectance material (Peplowski et al., 2016). Experiments on materials analogous to Mercury's mantle indicate that the only major mineral that would be buoyant in a Mercury magma ocean would be graphite (Vander Kaaden and McCubbin, 2015). Peplowski et al. (2016) inferred that the LRM in particular, and the volcanic upper crust generally, samples remnants of an early, graphite-rich flotation crust subsequently mixed and modified by impacts and magmatic intrusions and later excavated by large craters and/or assimilated into later volcanic magmas (Chapter 6). Attribution of the low reflectance of Mercury to elevated elemental carbon is consistent with the high S/Si ratio in surface materials. While graphite is a common mineral in enstatite chondrites, those meteorites contain only ~10% the C measured in CI chondrites.

### 18.2.5 Summary

The anomalously large Si-bearing Fe core, oxidation state, volatile enrichment, and reflectance of Mercury all demand explanation, but they are also clues to Mercury's formation. The Si enrichment of the core follows from the observed oxidation state, which probably also controls S distribution in the planet. In the meteorite record there exist rare materials that are similarly iron-rich, or similarly reduced, but not both. The EH enstatite chondrites are reduced; enriched in Si relative to Mg, Al, and Ca; and similarly volatile-rich. Although the EH chondrites contain Cl and K in high abundance, inferences from MESSENGER data for Mercury may represent lower bounds, subject to further experimental partitioning studies (McCubbin et al., 2012). The low reflectance of Mercury's surface may be closely related to its reduced chemistry if the planet formed with a substantial carbon content, as suggested by MESSENGER measurements (Peplowski et al., 2016).

## 18.3 Planet formation in disks

### 18.3.1 Theoretical considerations

The eighteenth century concept of the solar nebula (Kant, 1755) has evolved into modern astrophysical disk theory that treats three main stages of planet formation: (1) mineral dust concentrates in the midplane of the solar nebula and accretes to form multi-kilometer-sized planetesimals, (2) the largest planetesimals grow in annuli by runaway and preferential accretion to the largest bodies (oligarchic growth) to form Moon- to Mars-mass planetary embryos, and (3) final terrestrial planets form by energetic, stochastic collisions between embryos driven by gravitational interactions (Safronov, 1972; Morbidelli et al., 2012). Stage 1 is poorly understood and is thought to be rapid ($\sim 10^5$ yr), stage 2 ($\sim 10^6$ yr) forms embryos with characteristic spacing, with gas dissipating in a few million years, and stage 3 ($\sim 10^8$ yr) establishes the final radius, mass, and composition of each of the terrestrial planets through violent collisions between planetesimals and embryos from disparate solar radii (Chambers, 2004, 2009a). The challenge lies in understanding the details of how these processes occurred.

The cornerstone for chemical models for planet formation is the chondritic model for condensates in the solar nebula. The variations in the volatile compositions of the classes of chondritic meteorites are thought to reflect differences in the pressure–temperature conditions of equilibration of component solids with the gas in the solar nebula, overprinted with variable abundances of volatile ices (Figure 18.4; Davis, 2006). How these conditions varied over time with distance from the Sun and height in the disk is not known, but the conditions are probably recorded in the oldest, most volatile-rich meteorites, from bodies that did not differentiate into cores and mantles (Alexander et al., 2001). These meteorites come from asteroids and were accreted from high-temperature, 100–1000-micrometer-size chondrules and Ca-, Al-rich inclusions that were once free-floating nebular solids. In the chondrites least altered by water and heat on their parent bodies, these objects are surrounded by a fine-grained matrix containing presolar, interstellar grains; organic matter; amorphous particles; and other materials (Alexander et al., 2007). The physical origin of meteoritic assemblages of chondrules and matrix materials remains elusive (e.g., Ebel et al., 2016). Although separation of metal and silicate among different components is observable at the millimeter scale, and there is significant variation in the metal content of chondrite groups, only one group (CH chondrites) has a metal/silicate ratio similar to that of Mercury (Figure 18.3). Understanding the origin of Mercury is significantly handicapped by the lack of certainty about the formation of terrestrial planets in general. The physics of growth from dust to Mars-mass bodies is particularly poorly constrained. As a result, the chemical evolution of the precursor materials of planets cannot be robustly predicted, and certainly not as a function of time and solar distance.

In the past several years, new ideas have challenged traditional models of the orderly growth of planets. For example, planetesimals may rapidly grow into embryos by so-called "pebble accretion." In this model, the accretion efficiency of centimeter- to sub-meter-sized "pebbles" is greatly enhanced by Stokes drag in the atmospheres around growing embryos (Lambrechts and Johansen, 2012; Johansen et al., 2014). Calculations of embryo growth by pebble accretion have successfully led to systems that resemble the Solar System's outer planets (e.g., Chambers, 2014; Levison et al., 2015a) and the terrestrial planets (Levison et al., 2015b). However, neither the physical origin nor the chemical nature of pebbles is constrained because the models require only that pebbles are objects with a favorable Stokes number, when the gas-drag stopping time is comparable to the time it takes for the pebble to cross the embryo's region of gravitational influence (Hill radius). In most models, pebbles are larger than the mostly sub-millimeter-sized chondrules found in meteorites (Friedrich et al., 2015), which would be strongly coupled to the gas. Thus the relationship between pebbles and meteorites is currently unknown, and the chemical relationship between meteorites and growing embryos in this model is unexplored.

Central to the accretion of the terrestrial planets are the motions of the giant planets. The Nice model describing the outward migration of the giant planets from an earlier, more compact configuration is now a widely accepted basis for scenarios describing the early history of our Solar System (e.g., Tsiganis et al., 2005; Levison et al., 2007; Morbidelli et al., 2007; Batygin and Brown, 2010). The possibility of inward and outward migration, as proposed in the Grand Tack model (Walsh et al., 2011), is still under scrutiny. Competing models, such as pebble accretion, offer alternative solutions to the low mass of Mars (Levison et al., 2015b). Giant planet migration excites the orbits of bodies in the terrestrial zone, increasing the distribution of collisional energy and radial mixing of materials. Importantly, the probability of collisions of different energy depends on the overall context for terrestrial planet formation (e.g., Carter et al. 2015). For example, more destructive collisions are probable when the motion of a giant planet excites the orbits of bodies in the terrestrial region. By comparison, collisions generally lead to partial accretion during most of planet formation under standard models.

Most physical models of planet formation have not been well coupled to chemical models. However, recent research provides constraints on the chemical evolution of planets during the late stages of planet formation. Abundant evidence, including Earth's chondritic mantle $^{107}$Ag/$^{109}$Ag ratio (Schönbächler et al., 2010), correlated volatile–refractory element pair variations (Wänke, 1981), and young U-Pb ages of old feldspars (Albarède, 2009), suggests that volatiles must be delivered to Earth late or after the origin of the Moon. Other recent studies, however, have argued that Earth must have accreted water and other volatiles earlier in its growth history (Halliday, 2013; Dauphas and Morbidelli, 2014). Without a better understanding of volatile incorporation into all planets, it is difficult to assess whether the highly reduced state of Mercury falls in the expected range of final planets or if it requires an anomalous event that altered the chemical evolution of the planet.

Perhaps the only commonalities across all terrestrial planet formation models are the initial existence of a dust-rich nebula and a terminal period of giant impacts among planetary embryos. During the period of giant impacts, substantial chemical modifications of planets are possible (Wetherill, 1994; Asphaug, 2010; Stewart and Leinhardt, 2012; Asphaug and Reufer, 2014), including highly energetic collisions that could remove mantle material. The likelihood of mantle-stripping event(s) on a proto-Mercury can be assessed only in the context of specific models for the general process of terrestrial planet formation. However, dynamical studies of orbital evolution and merging of bodies leading to terrestrial planet formation (N-body simulations) universally exclude the inner region of the disk where Mercury orbits today because calculating the direct gravitational forces between all protoplanets on short orbital timescales is computationally expensive and difficult to parallelize.

Now that MESSENGER has provided new and important chemical observations, Mercury can be used in future studies as a powerful constraint and test of different models for terrestrial planet formation.

### 18.3.2 Observations

Protoplanetary disks are rotationally supported structures of gas and dust around young stars (Williams and Cieza, 2011; Armitage, 2011). Disks are observed around many T-Tauri type stars, actively accreting low-mass ($< 3M_{sun}$, where $M_{sun}$ is the solar mass) pre-main-sequence stars such the young Sun prior to initiation of nuclear fusion (McClure et al., 2013). Statistical analyses of populations indicate that protoplanetary disks persist only for a few million years (Haisch et al., 2001). Understanding the mechanism of the observed rapid spin down of low-mass stars to order 10% of the break-up velocity while they are actively but episodically accreting mass is a difficult astrophysical problem, since angular momentum transfer into the disk must be rapid (Hartmann, 2009). High accretion rates are associated with strong outflows (Reipurth and Bally, 2001) and X-ray emission (Feigelson, 2010). Strong and variable magnetic fields in such high-energy environments complicate magnetohydrodynamic (MHD) modeling (McNally et al., 2013). Yet it is during this stage that planetesimals grow, with chemical compositions that may be strongly affected by the local physical environment. If Mercury's anomalous composition results from early chemical–physical processes at solar distances less than 0.5 AU, understanding those processes presents an extreme challenge for both astronomical observation and astrophysical MHD models.

The conditions in the innermost regions of planetary systems have been further complicated by observations of extrasolar planetary systems. In some exosystems, the inner regions contain substantially more mass than our Solar System (Barnes et al., 2008; Fabrycky et al., 2014). For example, Kepler-11 has six known planets less than 5 Earth radii in size all orbiting inside of 0.5 AU (Figure 18.5; Lissauer et al., 2011). Closely packed inner planetary systems are not universal, however, and other systems are more similar to our own, with only one planet having an orbit less than 200 days (Fang and Margot, 2012). The bulk densities of close-in exoplanets are difficult to measure (Chen and Kipping, 2016). So far, observations are consistent with many rocky exoplanets having Earth-like metal core fractions (Dressing and Charbonneau, 2015; Zeng et al., 2016). Mass-radius relations for close-in exoplanets (Figure 18.5; Chen and Kipping, 2016) yield bulk densities exceeding that of Mercury. Future observations should enable better calculations of masses and radii for planets very close to their host stars. At this time, the abundance of dense, inner members of stable multi-planet systems (e.g., around Kepler-37, Barclay et al., 2013) suggests that Mercury analogs are not rare.

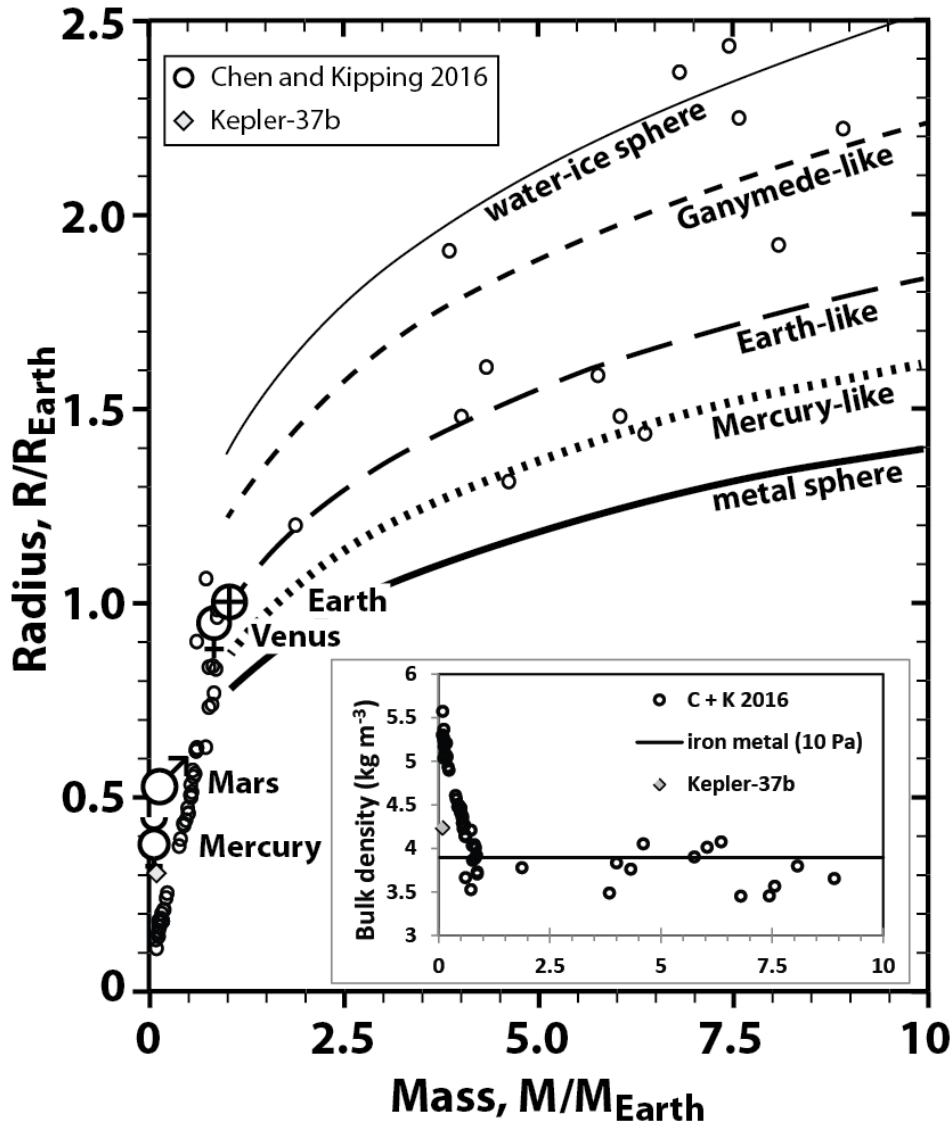

**Figure 18.5**. Mass-radius relations of selected low-mass (mass $M_p < 10\ M_{Earth}$), small (radius $R/R_{Earth} < 2.5$) exoplanets (Chen and Kipping, 2016, their Appendix I) compared with the terrestrial planets. Inset shows calculated bulk densities versus $M_p/M_{Earth}$ for the same data set. Curves are for Earth-like, Mercury-like, and Ganymede-like planets, and for a metal sphere and water-ice sphere (after Wagner et al., 2011). Values for Kepler-37b are from Barclay et al. (2013), using their relation $(M_p/M_{Earth} = (R_p/R_{Earth})^{2.05}$ and inferred radius. The tight, nearly linear behavior of the relations for the smallest exoplanets is a direct result of model assumptions.

## 18.4 "Chaotic" models for Mercury's origin

### 18.4.1 Modeling collisions

Several proposals for Mercury's origin are variations on removal or separation of material by collisions. Because the possible scales of impacts vary from numerous small impactors to a single giant impact event, the possible time frame for the origin of Mercury's mass anomaly could span the period from embryo growth to the final stages of planet formation. Until recently, numerical *N*-body simulations of terrestrial planet formation were based on the assumption that collisions between bodies resulted in perfect merging (e.g., Kokubo and Ida, 1996, 1998; Raymond et al., 2009). As a result, these studies could not directly address the question of removal of silicate material by energetic collisions.

Major developments have improved numerical models of compositional evolution during planetary accretion. First, new analytic formulations have simplified the calculation of collision outcomes (Leinhardt and Stewart, 2011; Genda et al., 2011; Leinhardt et al. 2015). These formulations have been partially implemented into *N*-body simulations (Chambers, 2013; Dwyer et al., 2015; Bonsor et al., 2015; Carter et al., 2015; Leinhardt et al., 2015; Quintana et al., 2016). In general, the number of collision fragments is limited in order to keep the calculation tractable, a restriction that prevents detailed investigation of processes that involve the smallest bodies. Only one of these studies explicitly tracked the evolving metal mass fraction of the bodies and debris (Carter et al., 2015), and most studies have attempted to investigate compositional variations by post-processing the simulation data, although this approach is not robust (Bonsor et al., 2015). Carter et al. (2015) focused on growing planetary embryos; this stage did produce diversity in metal/silicate mass ratios of embryos and planetesimals, with greater variations in the migrating giant planet Grand Tack scenario. However, embryos with metal mass fractions as high as Mercury were not produced in the limited number of simulations.

Second, *N*-body simulations with perfect merging after collision have been combined with metal–silicate equilibration models to predict the evolution of core and mantle compositions during accretion (e.g., Rubie et al., 2015). Such studies can address the redox state of the mantle of a growing planet subject to assumptions about the initial composition and chemistry of the embryos and planetesimals and how the initial ice/rock ratio varies through the Solar System. The evolution of the redox state of accreting planets is very much a topic of debate. In the approach of Rubie et al. (2015), variations in oxidation state are primarily controlled by the addition of water ice. In alternative approaches that are not fully coupled to accretion simulations (e.g., Wood et al., 2006; Badro et al., 2015), the mantle oxidation state evolves primarily from changes in the pressure–temperature conditions on the planet rather than from changes in accreting material. The development of linked accretion and compositional evolution models may be able to address the origin of Mercury's reduced mantle.

Third, our understanding of individual giant impacts is largely guided by studies of the origin of the Moon. Over the past several years, the giant impact hypothesis for lunar origin has been scrutinized (Asphaug, 2014; Melosh, 2014) because of the conflict between the predicted composition of the Moon and observations that the isotopic compositions of Earth and the Moon are very similar (Burkhardt, 2014). The canonical model (Canup and Asphaug, 2001; Canup, 2004, 2008) predicts that most of the lunar material would be derived from the impactor, which is expected to have isotopic signatures different from those of the proto-Earth. This conflict motivated studies of new styles of giant impacts that predict similar fractions of impactor material in Earth and the Moon (Canup, 2012; Ćuk and Stewart, 2012); however, these proposed solutions did not fully resolve the similarities and differences in the chemical makeup of Earth and the Moon (Elkins-Tanton, 2013; Asphaug, 2014; Melosh, 2014). The latest developments in lunar origin studies are predictions of the details of the chemical composition of the Moon due to its formation in a circumterrestrial disk (Canup et al., 2015; Lock et al., 2016). These studies find that the depletion of volatile elements in the Moon is due to incomplete condensation in the circumterrestrial disk rather than thermally driven loss by the energy of the giant impact. This result has direct implications for proposed collisional models for the origin of Mercury.

**18.4.2 A giant impact**

A longstanding proposal for the origin of Mercury's large core is stripping of most of the planet's mantle by one or more giant impacts (Smith, 1979; Benz et al., 1988; Cameron et al., 1988). This scenario typically involves a differentiated proto-Mercury with more than twice the present mass that is impacted by a smaller differentiated body with sufficient energy to disperse a portion of the silicate mantles of the two bodies. If the colliding bodies had initial core mass fractions similar to that of Earth, a single impact must gravitationally disperse about half the total mass, primarily the silicates, to achieve Mercury's current core mass fraction (Benz et al., 1988, 2007). This class of collisions, called "catastrophic," is achieved by impact energies equivalent to a few to several times the gravitational binding energy of the total mass (Leinhardt and Stewart, 2011). Thus, most of the impact energy is converted into thermal energy, and a substantial portion of the planet and ejected fragments are transiently vaporized (Benz et al., 2007).

In a catastrophic collision between differentiated bodies, most of the ejecta are derived from the silicate layers of the colliding bodies, and most of the metal cores merge to form the new core of the largest post-collision body (Marcus et al., 2009, 2010b). The total mass of the largest remnant is determined by the self-gravity of a transiently decompressed, hot cloud of (generally segregated) metal and silicate debris, and the timescale of re-accretion into a compressed planetary structure is several hours. At that point, the planet would have a gravitationally separated molten core and silicate mantle. The mantle would have gained sufficient entropy to be a liquid to supercritical fluid at the highest pressures and pure vapor at the lowest pressures. At the lowest pressures, metal may be miscible in the silicate fluid. Mercury's thermal state would be analogous to that of Earth after the Moon-forming giant impact (e.g., Canup, 2008; Nakajima and Stevenson, 2015).

While the dynamics of catastrophic impacts are reasonably well understood, the potential for chemical changes to the colliding bodies are not. For example, the extent of metal–silicate re-equilibration is not understood in giant impacts because numerical simulations cannot model the small-length-scale processes that would enable greater mixing and chemical equilibration. Incomplete metal–silicate equilibration during accretion is inferred for Earth on the basis of the $^{182}$W content of the mantle (Kleine et al., 2004; Rudge et al., 2010; Rizo et al, 2016). From dynamical considerations, portions of the cores of the impacting objects directly merge during giant impacts (Marcus et al., 2009), and only a portion of the silicate mantle is equilibrated with each small impactor (Dahl and Stevenson, 2010; Morishima et al., 2013). Thus, the effects of a giant impact on the redox state of Mercury's mantle and core, and their degree of subsequent equilibration, remain unquantified.

One of the major questions about the proposed giant impact origin for Mercury is its effect on the volatile content of the planet. When MESSENGER determined that the K/Th and K/U ratios of Mercury are similar to those of the other terrestrial planets, the giant impact hypothesis was rejected as inconsistent with a high-temperature event (Peplowski et al., 2011). At the time, there were no quantitative models available to address how moderately volatile elements would be affected by a giant impact. As a result, the observed depletion in volatile and moderately volatile elements in the Moon (Ringwood and Kesson, 1977) guided inferences about the outcome of giant impact events.

However, all models of planet formation predict that the terrestrial planets (with the possible exception of Mars) experienced a stochastic number of giant impact events during accretion.

Thus, if giant impacts substantially depleted moderately volatile element abundances on growing planets by removal, one would expect a different magnitude of depletion on each planet. The similar K/Th and K/U ratios for terrestrial planets, for which the impact histories must vary greatly, imply that giant impacts are not primarily responsible for the magnitude of moderately volatile element depletion in planets compared with CI chondrites (Figure 8.4), although detailed chemical calculations have not been accomplished to address this process. Giant impacts primarily remove bulk mantle silicates without chemical fractionation. Vaporized mantle would remain gravitationally bound and would recondense upon planetary cooling (Stewart et al., 2013, 2016).

This inference is supported by recent calculations demonstrating that the depletion of K and Na on the Moon compared with Earth is a result of incomplete condensation in the lunar disk (Canup et al., 2015; Lock et al., 2016). Hence, the abundance of moderately volatile elements on the Moon is different from those of the planets because of the Moon's origin in a circumplanetary disk (Stewart et al., 2016). Instead, Earth's moderately volatile element abundances are a better analog to a post-impact Mercury.

After a giant impact, a planet is a mixture of the compositions of the colliding bodies, with the portion from each body strongly dependent on the impact geometry and velocity (e.g., Canup, 2004, 2012; Ćuk and Stewart, 2012; Reufer et al., 2012). Because the post-impact planet is defined by self-gravitationally bound material, thermally driven escape is limited to the high-velocity tail of the Boltzmann distribution and the small mass fraction in the collisionless outer regions of the atmosphere. Radiative cooling quickly leads to re-condensation of silicates and sulfides from the vapor atmosphere, and thermally driven escape is limited (Stewart et al., 2016). At present, a giant impact origin for Mercury does not have a predicted chemical signature, other than the motivating observation of an enhanced metal core fraction, with which to test the hypothesis.

Even in the absence of a chemical test for the giant impact hypothesis, such an origin for Mercury is difficult. The principal problem with a single impact hypothesis is the re-accretion of debris. The orbits of gravitationally ejected material intersect Mercury's orbit. Numerical simulations that track debris production and re-accretion find that most of the debris is quickly re-accreted, which limits the variation in core fraction of growing embryos (Carter et al., 2015).

Benz et al. (1988, 2007) proposed that the ejecta could be separated from Mercury if the ejected particles were sufficiently small to evolve quickly under non-gravitational forces such as Poynting-Robertson drag. Benz et al. (2007) calculated that most of the ejecta would be shock-heated to vapor and recondense as ~centimeter-sized particles. They calculated the orbital evolution of the ejecta and found that about one-third was re-accreted to Mercury in the first 2 Myr. This timescale is comparable to the half-life of centimeter-sized particles undergoing Poynting-Robertson drag into the Sun. Benz et al. (2007) proposed that Mercury could be formed by a "super-catastrophic" impact onto proto-Mercury, followed by partial re-accretion of the ejecta and substantial loss of ejecta to the Sun.

Gladman and Coffey (2009) also considered the dynamics of ejecta from a giant impact onto Mercury. They found that the debris field would be optically thick, which limits the role of Poynting-Robertson drag. In addition, the debris would collide with itself, further increasing the optical thickness and likelihood of re-accretion onto Mercury.

Changes in the chemical composition of the ejecta have not been modeled in detail. If the ejecta cool sufficiently quickly so that silicates and sulfides recondense faster than the timescale of dynamical separation of condensates and gas, then the composition of re-accreting ejecta may be similar to that of the original colliding planets. As mentioned above, the dynamical evolution and re-accretion of condensed ejecta have timescales of millions of years (Benz et al., 2007; Gladman and Coffey, 2009). Adiabatic expansion and radiative cooling would lead to re-condensation of silicates and sulfides on much shorter timescales (e.g., days to years). Whereas highly volatile elements may remain in the gas phase in orbit around the Sun, the condensed ejecta should have compositions similar to the silicate layers of the source planets.

Ejecta may also be accreted onto other planets or planetary embryos. If some other nearby planet had a larger mass than Mercury, a majority of ejecta could be accreted to that body. Such a scenario has motivated a variation of the giant impact hypothesis for Mercury in which proto-Mercury is the smaller body in a so-called "hit-and-run" event.

### 18.4.3 A hit-and-run impact

Agnor and Asphaug (2004) first realized that a large fraction of collisions between similarly sized bodies do not lead to mergers. During the giant impact stage (3), about one-third of collisions between planetary embryos are hit-and-run events where the two bodies obliquely collide and then gravitationally separate (Stewart and Leinhardt, 2012). In most cases, the two bodies collide and merge in a subsequent encounter (Kokubo and Genda, 2010; Carter et al., 2015). However, it is possible for a planetary embryo to be scattered to the inner edge of the terrestrial accretion zone and survive to the end of accretion (Hansen, 2009).

In a hit-and-run encounter, the smaller body may be catastrophically disrupted (Asphaug et al., 2006; Asphaug, 2010; Leinhardt and Stewart, 2011). Sarid et al. (2014) and Asphaug and Reufer (2014) proposed that proto-Mercury was the smaller body. Stripping the mantle of the smaller body requires a less energetic collision – or multiple low-energy collisions – than in the standard single giant impact scenario. However, the arguments presented above regarding the lack of chemical changes (e.g., volatile content and redox state) to the silicate mantle would also apply to the hit-and-run scenario regardless of the intensity of heating and vaporization.

In the hit-and-run variation, the ejecta would primarily be accreted onto the larger-mass body, perhaps a proto-Venus planetary embryo. Multiple hit-and-run events are possible between a pair of embryos, so this variation does not require that the enhanced metal mass fraction of Mercury be achieved in a single event. At this time, hit-and-run scenarios have not been tested in full numerical simulations of terrestrial planet formation, let alone with chemistry, so the likelihood of a Mercury-like outcome is not known. The key issue for this hypothesis is preventing the smaller proto-Mercury from ultimately being accreted onto the larger body.

The amount of mass lost from the smaller body is extremely sensitive to the impact geometry and velocity. At this time, the numerical simulations of individual hit-and-run encounters have not yet been distilled into a simple analytic formula, which is needed for future investigation of this scenario in *N*-body simulations of planet formation.

### 18.4.4 Collisional erosion

Another variation on collisional stripping of Mercury's mantle is the accumulated effects of many small high-velocity impactors (Vityazev et al., 1988; Svetsov, 2011). In these scenarios, the impact velocities must exceed ~25 km/s in order for each collision to eject a total mass greater than that of the impacting body (Svetsov, 2011). For example, in order to erode a proto-Mercury with an Earth-like core fraction to the present core mass fraction requires more than a Mercury mass of planetesimals impacting at ~30 km/s (Svetsov, 2011). During planet formation, the typical collision velocities are a factor of 1 to 3 greater than the escape velocity of the largest bodies (e.g., O'Brien et al., 2006; Raymond et al., 2009). Thus typical Mars-mass embryos would not experience primarily erosive planetesimal bombardment. Indeed, most planetesimal collisions result in net accretion unless a migrating giant planet dynamically excites the system (Carter et al., 2015). Even if such a dynamically excited bombardment occurred, the re-accretion of ejecta remains an issue unless ejecta could be preferentially accreted onto other protoplanets. In addition, heavy bombardment of Mercury would also affect the composition of the other inner planets.

The investigations of collisional erosion of Mercury illustrate the general problem with invoking collisional erosion of planetary crusts to account for geochemical observations, as first proposed by O'Neill and Palme (2008). The required high-velocity bombardment is not predicted in current models of planet formation. In a giant impact, portions of both mantle and crust are ejected rather than all of the crust. Numerical simulations that include probable collision outcomes indicate that small-scale debris is continuously recycled as it is ejected and then re-accreted during the growth of planetary embryos (Bonsor et al., 2015; Carter et al., 2015). Thus, collisions are not expected to permanently remove incompatible elements that are concentrated in planetary crusts. Indeed, Mercury does not appear to have experienced such removal.

Bulk ejection of silicate layers, and their likely re-accretion, would not chemically fractionate the silicate component of the planet. Thus, the observed chondritic Cl/K ratio (Evans et al., 2015) on Mercury does not provide a major constraint on the impact history of the planet. Venus, Earth, and the Moon have lower bulk Cl/K ratios than Mercury, Mars, and chondrites (see Evans et al., 2015). The origin of these differences among bodies is not likely to be a result of collisional erosion of planetary crusts. Indeed, the role of collisional erosion in the inner Solar System was originally motivated by Earth's $^{142}$Nd abundances (Boyet and Carlson, 2005). These data have recently been reinterpreted such that no geochemical support for collisional erosion remains (Bouvier and Boyet, 2016; Burkhardt et al., 2016).

## 18.5 "Orderly" processes for Mercury's origin

The arguments for a "chaotic" stage of planet formation (stage 3) after the formation of embryos (stage 2) are strong. Such embryos would reflect the chemical conditions and dynamical processes in narrow radial annuli or feeding zones. The chemical signatures of planetesimals and embryos established during stage-2 formation would be largely scrambled during stage 3, remaining only as cryptic heterogeneities such as perhaps the oxygen isotopic differences among Earth+Moon, Mars, and Vesta (Clayton and Mayeda, 1996). However, Mercury could represent an embryo, a remnant of stage 2, and record in its chemical composition extreme processes occurring uniquely in the innermost solar nebula. Several such extreme processes have been put forward to explain Mercury's anomalous density. Here we consider each in the light of MESSENGER results.

### 18.5.1 "Old school" condensation

Following Urey (1950), Lewis (1973) stated that "the present quantitative theory for the composition and volatile content of Solar System bodies attributes both to one parameter alone, "formation temperature" (cf. Lewis, 1972, 1988). In this scenario, a hot disk of dust and gas, with dust concentrated in the midplane and innermost few astronomical units, cools and condenses solids slowly over time from the outer material toward the center (Larimer and Anders, 1967; Grossman and Larimer, 1974; Ebel, 2006), and those solids accrete onto the planets as we see them today, with very limited mixing. The best 20th-century estimates for the bulk compositions of Venus and Mercury were derived with this assumption (Morgan and Anders, 1980; Lodders and Fegley, 1998). Modern theories of disk structure and evolution have gone well beyond such a simple picture, which is inconsistent with both the observed architectures of exoplanetary systems and the volatile enrichment of Mercury's silicate fraction discovered by MESSENGER. Furthermore, as discussed in detail by Weidenschilling (1978), a temperature at which the Fe/Si ratio of solid condensates matches Mercury's Fe/Si ratio is possible, but removal of nebular gas must occur within a very narrow cooling window of 10–50 K, before significant Mg-silicate condensation (cf. Ebel, 2006, Plates 7 and 10).

Cameron (1985) suggested that the early Sun was sufficiently hot that the mantle of differentiating Mercury evaporated and was stripped by strong solar wind. Fegley and Cameron (1987) calculated that vaporization of ~70–80% of the silicate from an initially chondritic mantle would be required to match Mercury's uncompressed density. This combined thermochemical/dynamical model predicts strong depletion in moderately volatile elements (K, Cl, S, Na) and enrichment in refractory lithophile elements (Ca, Al, Ti, Th). Both $SiO_2$ and FeO become depleted in the mantle, and U would deplete relative to Th due to its propensity to form the gaseous oxide. This model is not supported by the MESSENGER results.

Morgan and Anders (1980) superposed physical processes on the original nebular condensates, such as preferential settling, size sorting, or ferromagnetic attraction, as well as thermal processing (melting, evaporation). Condensation arguments and observed K and U abundances were used to derive planetary compositions (Figure 18.3); however, they chose to use lunar values in the absence of data from Mercury. Going further, they assumed, "tenuously!" in their words, a mantle FeO abundance trend for the terrestrial planets that increases monotonically with solar distance, with Mercury pyroxenes at 5.5 wt% FeO. Results from MESSENGER invalidate such assumptions.

## 18.5.2 Metal/silicate fractionation

Chondritic meteorites represent the oldest, undifferentiated materials formed in the Solar System and are thought to be the precursors of the terrestrial planets. Metal abundance is one of three principal components describing the variation among chondritic meteorites (Grossman, 1988). There are numerous hints in primitive chondrites of metal fractionation from silicates early in Solar System history. Rare, metal-rich chondrites may record metal-silicate fractionation in some part of the early nebula. Iron-rich asteroids also suggest metal-silicate fractionation by dynamical processes.

### 18.5.2.1 Metal-rich chondrites

The metal-rich Bencubbin-like (CB) and CH carbonaceous chondrites (Weisberg et al., 2001) have many primitive characteristics and a higher grain density, due to an abundance of metal grains, than uncompressed Mercury (Figure 18.1). Chondrules and metal grains in these chondrites have been attributed to planetesimal impact up to 5 Myr after CAI formation, the usually accepted "origin" time of the Solar System (Krot et al., 2005; Fedkin et al., 2015), although some of their compositional characteristics are difficult to reconcile with such an origin (e.g., their CAIs, Weisberg et al., 2016). Collisions among bodies of ordinary chondrite composition in a disk from which most of the $H_2$-rich gas has been removed will recondense solids from oxidizing vapors, yielding FeO-rich olivine. Carbonaceous chondrite vapor is even more oxidizing (Ebel, 2001), so persistence of $H_2$-rich gas in the solar nebula is required to make these metal-rich meteorites in late-stage impacts. Persistence of a gas is also required to prevent fragmentation during chondrite accretion. These metal-rich chondrites, however, are not nearly as reduced as the enstatite chondrites, which contain CaS and MgS. If the CB chondrites formed late in Solar System history, their influence on planetary embryo compositions is likely to have been minimal.

### 18.5.2.2 Asteroid cores

A large number of magmatic iron meteorite parent bodies are inferred from the meteorite record, but complementary planetesimal mantles are missing. Differential comminution by impacts, followed by Poynting-Robertson drag over long time periods, is the most likely explanation for loss of silicates, primarily to the Sun (Burbine et al., 1996). W-Hf isotopic chronometry shows that the differentiation of many parent bodies occurred within 1 Myr of CAI formation (Yang et al., 2007; Burkhardt et al., 2008; Kleine et al., 2009). Density determinations appear to require that many iron-rich asteroids are bound, macroscopic, metallic rubble (e.g., 216 Kleopatra, Ostro et al., 2000; Consolmagno and Britt, 2004; Carry, 2012). The behaviors of big metal chunks and small, brittle mantle silicate fragments in the disk before and after gas dispersal would differ (Weidenschilling, 1978; section 18.5.2.3). However, the iron-rich asteroids represented in the meteorite record do not offer themselves as potential sources for Mercury's excess metal, since they are, with few exceptions, lacking in the alloyed metallic Si typical of highly reduced bodies (Section 18.2.2). The oxidation states of iron meteorites are quite different from that inferred for Mercury, on the basis of MESSENGER observations and experiments (Section 18.2).

### 18.5.2.3 Photophoresis and dynamical fractionation

The photophoretic effect has been explored to explain the wide variation in chondrite meteorite metal contents (Wurm et al., 2013) and size sorting of particles (Loesche et al., 2016). Photophoresis occurs when non-isotropic (e.g., Sun-derived) thermal radiation produces thermal gradients in millimeter-sized particles that drive those particles toward colder regions. This force is proportional to the temperature gradient in each particle (Krauss and Wurm, 2005). Differences in radial velocity relative to the Sun could separate metal, rock, and dust aggregate particles because metal has much higher thermal conductivity, so metal grains lack temperature gradients. Thus, photophoresis could have affected early collisional dust evolution and accretion (Loesche et al., 2016). If such effects were widespread in a particular annulus of the disk, resulting planetesimals could have inherited the effects of photophoresis, including metal/silicate fractionation (Cuzzi et al., 2008). This effect is expected to be strongest closer to the Sun.

Opacity is expected to be high in the midplane of a planet-forming disk, inhibiting the photophoretic effect. For photophoresis to work, a sufficient fraction of Mercury precursor material must see sufficient sunlight, and rocky grains must be forced outward, leaving an anomalously metal-rich feeding zone for Mercury. However, inner disk "walls" are observed in transition disks, which are stellar objects with substantially cleared inner disks, perhaps the result of planet formation (Espaillat et al., 2014), suggesting that accretion zones closest to stars could be affected by photophoretic separation of grains on the basis of their physical properties, e.g., size and porosity (McNally and McClure, 2017). Testing this hypothesis requires a model that addresses the chemistry of small solids in the inner disk, subject to photophoresis, and their accretion onto planetary precursors.

An "aerodynamic fractionation" model was proposed by Weidenschilling (1978), who recognized that more effective removal of silicates relative to metal from a chondritic feeding zone could account for Mercury's anomalous density. In the inner disk, small grains were entrained in the accreting gas, whereas larger grains (>1 m) in Keplerian orbits experienced a headwind due to the radiation pressure experienced by the gas. This gas drag caused meter-sized grains to drift sunward faster than the gas. Larger and/or denser boulders experienced slower orbital decay. Weidenschilling proposed that precursor multi-meter-size solids (boulders) with differing Fe/Si ratios could dynamically interact such that silicates were preferentially removed into the Sun by gas drag.

Hubbard (2014) explored a "magnetic erosion" model that combined the magnetic attraction of metallic grains and the differential comminution of silicates and metal by collision. He called on particular conditions at the outer edge of the inner disk to enhance magnetization of metal-rich grains, thus enhancing both the collisional removal of non-magnetic silicate and the rapid collisional growth of large metal grains to sizes beyond the meter barrier. The magnetic field requirement restricts this mechanism to the inner disk.

All three of these models are consistent with Mercury's anomalous density and Earth-like mantle abundances, and they all enrich Mercury in metal that is commonly associated with sulfide, consistent with Mercury's likely S enrichment. None of these models, however, addresses the extremely reduced nature of Mercury. They all call on special astrophysical conditions in the innermost nebula, and none has been explored with fully three-dimensional chemical–physical models. If precursor metal is assumed to carry sulfur, then Mg-silicates and refractory elements might be assumed to be in large grains (chondrules, CAIs), whereas K, Na, and Cl would have concentrated in smaller grains, the matrix mineral dust of chondrites. MESSENGER results for K indicate that Mercury's volatility curve is not steeper than Earth's (Figure 18.4). The fates of the volatiles in these scenarios are difficult to predict.

Several recent astrophysical models have yielded results consistent with Mercury's metal enrichment and reduced chemistry for disks around stars similar to the Sun. Pasek et al. (2005) used a two-dimensional steady-state α-disk solution to yield time-temperature-pressure histories in the disk midplane at various solar radii along with chemistry codes to compute condensation fronts. With a model for diffusive transport driven by condensation fronts in the inner disk, they then calculated condensation histories at progressively more O-depleted inner radii. They found that the effects of water depletion on nebular S speciation formed reduced enstatite-chondrite-like rocks at Mercury-like astrocentric radii.

Moriarty et al. (2014) applied a disk model (Chambers, 2009b) to calculate disk temperature, pressure, and density over time. They then calculated equilibrium chemistry at steps in time and radius, removing gas and dust into planetesimals growing at a prescribed rate and decoupled from the gas. They also accounted for the effects of radial gas movement (with perfectly coupled dust) on the chemical inventory. Finally, their planetesimals were input into an N-body simulation of late-stage planet formation. Their model produced C-rich, short-period planetesimals around stars with C/O ratios slightly above the solar ratio.

Pignatale et al. (2016) used a disk model (D'Alessio et al., 1999) to prescribe temperature and pressure in a two-dimensional disk and then calculated condensation under each of those conditions. They then applied dust-settling and radial-migration models to calculate redistributions of material. Their model produced sulfide- and enstatite-rich zones within 1 AU of the young Sun. Although feedback between chemistry and dynamics is highly limited or missing from efforts to date to couple chemistry and dynamics, model results consistently predict reduced, metal-enriched inner disks, the probable feeding zone for planet Mercury.

### 18.5.2.4 C-rich condensation

The gross separation of inner volatile-poor planets and outer volatile-rich planets immediately suggests a "snow line," a complex time-dependent surface inside of which water remained in the vapor phase. In the presence of free oxygen, a similar "C-line" would have marked the locus inside of which graphite would have oxidized to CO and $CO_2$. The time–temperature and oxygen distribution history of the inner disk are not known, nor is the time-dependent accretion flux of interstellar material onto the innermost disk well quantified. The balance between C and O as functions of time and radius in the terrestrial planet-forming region may, therefore, have been quite heterogeneous. Ebel and Alexander (2011) investigated the consequences of carbon enrichment and oxygen depletion on the stability of minerals in a cooling $H_2$-rich vapor. They noted that the most abundant interplanetary dust particles (IDPs) in the present Solar System are 50–1000 μm sized, anhydrous, porous, chondritic "C-IDPs." The C-IDPs are aggregates containing highly primitive submicrometer silicates, metal, sulfide and presolar grains all attached together by poorly graphitized carbon (Messenger et al., 2003; Busemann et al., 2009; Bradley, 2014). Their original C content has been diminished by pre-capture stratospheric entry.

Ebel and Alexander (2011) explored the consequences of equilibrium condensation in systems enriched in a C-enriched, O-depleted analog C-IDP dust. At high (1000×) enrichments in such a dust (relative to $H_2$), condensates at 1650 K (at 10 Pa total pressure) have atomic Fe/Si reaching 50% of the estimate for bulk Mercury, because Si remains in the vapor to low temperatures (< 1000 K) as $SiS_{(gas)}$. Sulfur behaves as a more refractory element in this reducing system than in systems enriched in chondritic dust (Ebel and Grossman, 2000), stable minerals are FeO-poor, and the minerals stable at high temperature include CaS and MgS. These results reproduce mineralogical and petrological characteristics of enstatite chondrites. These calculations suggest that enrichment of an anhydrous chondritic dust, with variation in O/C ratios by O depletion, can produce environments in which highly reduced, high Fe/Si condensate assemblages are stable above 1200 K. If such condensates were isolated from the gas phase, they could have formed enstatite chondrite parent planetesimals, which could then have accumulated to form Mercury-like planets.

In follow-up work, Ebel and Sack (2013) computed the stability of djerfisherite, $K_6(Fe,Ni,Cu)_{25}S_{26}Cl$, in similar C-IDP dust-enriched systems. Comparison with djerfisherite occurrences in EH3 enstatite chondrites allowed them to conclude that both K and Cl, as well as S, can behave as refractory elements under the hypothesized nebular conditions.

This line of reasoning relies on the assumption that C-rich silicate dust could reach the inner Solar System well inside the evaporation radius of more oxidized material. There are other hints at such chemical gradients, but all are cryptic. Initial enrichment of Mercury in elemental carbon would provide both a light element in the core (in addition to S and Si) and an explanation for Mercury's low spectral reflectance (Vander Kaaden and McCubbin, 2015; Peplowski et al., 2016).

### 18.6 Bearing on extrasolar planets

We are entering the third decade of a revolution in the observation and theoretical understanding of extrasolar planets. One of the critical questions regards the relationship between observed high-mass inner planets, "super-Earths," and the terrestrial planets in our own

Solar System (Batalha et al., 2011; Morbidelli and Raymond, 2016). Wagner et al. (2011; cf. Dressing and Charbonneau, 2015; Zeng et al., 2016) calculated radii $R/R_E$ and masses $M/M_E$ for hypothetical, fully differentiated, thermally equilibrated Earth-like (32.5 wt% iron core, 67.5% silicate mantle), Mercury-like (70% core, 30% silicate), and Ganymede-like (6.5% core, 48.5% silicate, 45% water ice) planets (Figure 18.5), where $M_E$ is Earth's mass and $R_E$ is Earth's radius. These curves are illustrated in Figure 18.5 and compared with inferred properties of a subset of planets discovered up to early 2016.

The difficulty of estimating exoplanet mass and size is illustrated by the exoplanets that plot below the low-mass extension of the metal sphere mass-radius curve of Wagner et al. (2011) in Figure 18.5. The observational bias and the large uncertainties in data even for transiting planets make comparisons with our terrestrial planets premature. Marcus et al. (2010a) estimated the maximum amount of mantle material that could be removed from a planet by a single giant impact. This minimum radius is widely used as a lower limit on the radius of rocky exoplanets because pure-iron planets are unlikely. Planets with densities indicative of high metal core fractions are at the lower edge of the range of planet sizes and orbital periods explored by the Kepler mission (Borucki, 2016), but the abundance of Mercury analogs in the existing data (e.g., Barclay et al., 2013; Chen and Kipping, 2016) suggests that they are a natural outcome of dynamical–cosmochemical processes in protoplanetary disks.

## 18.7 Summary

### 18.7.1 "Chaotic" models

The origin of Mercury remains elusive. The MESSENGER mission has provided strong constraints on potential models, but models to date lack the detail to be tested against the observations. Here, we summarize the major models and their current challenges. Three dynamical-impact-related or "chaotic" models (section 18.2.1) have been considered:

- A single giant impact onto proto-Mercury: The energy required for a single event falls in the expected range of planet formation. The primary challenge lies in the expected re-accretion of debris. The orbits of the ejected silicates must be dynamically separated from the post-collision Mercury.
- A hit-and-run with proto-Mercury as the smaller body: This type of encounter is expected to be common during stage 3 of planet formation. Preferential accretion of debris onto the larger body is possible but has not yet been investigated quantitatively. The model demands that the post-impact Mercury not also be accreted onto the larger body, which may require special circumstances such as dynamical ejection out of the zone of planet formation.
- Multiple impact scenarios: Current models for terrestrial planet formation do not predict a bombardment of small planetesimals with sufficient energy to remove Mercury's mantle if it began with a chondritic bulk composition. A mixture of small and large impacts may produce a range of final core mass fractions, but the limited simulations conducted so far have not produced core enhancements as large as observed on Mercury.

All such models of the inner Solar System have fallen short of directly addressing the origin of Mercury. Future detailed calculations of terrestrial planet formation will be able to address the plausible range of final core mass fractions during collisional accretion and evolution of planets. Combined with chemical models of the growing planet, such studies will also be able to address the origin of Mercury's extremely reduced mantle and range of possible core compositions, constrained by experimental petrology. At present, there are no obvious geochemical tests of the collisional origin hypotheses. Current geochemical observations from all the terrestrial planets and physical studies of collisional accretion do not support earlier suggestions that collisions should preferentially remove moderately volatile elements or incompatible elements.

**18.7.2 "Orderly" models**

"Orderly" processes (Section 18.2.1) perhaps fare better than "chaotic processes" in approaching explanations for Mercury's origin. MESSENGER findings definitively refute previous models in which only the most refractory elements condense and accrete to form Mercury from a vapor of solar composition in the hot innermost part of the protoplanetary disk (e.g., Section 18.2.3; Morgan and Anders, 1980). MESSENGER results also refute early hypotheses of direct evaporation of silicates from proto-Mercury by an energetic young Sun. Such a process would rapidly deplete Mercury in many moderately volatile elements, including Si and Mg. The Si-poor metal in most metal-rich chondritic or iron-rich asteroidal material and the measurable FeO in silicates in most chondrites eliminate their potential role as major precursors to highly reduced Mercury. Only a very tiny fraction of Solar System materials in the meteorite record, the EH chondrites, have experienced reducing conditions similar to Mercury's.

Dynamical models in the context of a disk in which composition, temperature, and density vary smoothly with heliocentric radius offer another route to chemical fractionation close to the Sun. Although photophoresis models are not sufficiently mature to produce Mercury's density and redox anomalies, they do separate metal from silicate, particularly in the innermost disk. Aerodynamic fractionation of Fe-rich "boulders" from smaller, brittle silicates and preferential removal of non-magnetic silicates by "magnetic erosion" are both consistent with Mercury's anomalous density, but neither model appears to address the highly reduced chemistry of Mercury.

The possibility that the inner region of the disk was carbon-rich allows consideration of processes that could yield Mercury-producing scenarios. Condensation in C-rich, O-poor systems has been explored with very conservative parameters, yielding enstatite chondrite condensates (Ebel and Alexander, 2011). This parameter space might yield condensates with high Fe/Si at high temperatures. There is strong evidence for the volatility-controlled accretion of solids onto various chondrites and the planets. A scenario of high C/O condensation would require a dynamical argument for accretion of high Fe/Si solids at the high temperatures at which Si remains in the vapor phase. A successful model describing the physical chemistry of the inner disk must address radiative transfer, chemistry–opacity feedbacks, and active mass transfer among vapor, liquid, and solid phases.

**18.8 Conclusion**

MESSENGER confirmed the anomalous bulk Fe/Mg ratio and anomalously reduced nature of Mercury and demonstrated that the abundances of moderately volatile elements are consistent across the terrestrial planets and asteroids. These findings indicate that these bodies record disk processing rather than special circumstances of individual planets. Models that focus on terrestrial planet formation do not commonly produce Mercury-like planets. However, the EH chondrites offer a way forward in understanding how gradual processes in the innermost disk might have produced a small, reduced, Fe- and probably Si-rich planet.

It is plausible that the present state of Mercury is the sum of several processes at work during planet accretion. None of the proposed formation models can explain all of the observations, and all suffer from our lack of information about the chemical, thermal, and dynamical conditions at the inner edge of the protoplanetary disk. Increasingly detailed astronomical observations of protoplanetary disks and other planetary systems will aid development of more detailed planet formation theories. Such theories must be informed by better coupling of physical and geochemical processes in models of inner disk processes. Whether Mercury requires a special circumstance to account for its unique properties remains to be answered.

**Acknowledgements**

The authors thank Erik Asphaug, Larry Nittler, and Sean Solomon for comprehensive reviews. Denton Ebel thanks the Arthur Ross Foundation and American Museum of Natural History for support of his participation on the MESSENGER Science Team. This research has made use of NASA's Astrophysics Data System.

**References**

Agnor, C. and Asphaug, E. (2004). Accretion efficiency during planetary collisions. *Astrophys. J. Lett.*, **613**, L157–L160.

Albarède, F. (2009). Volatile accretion history of the terrestrial planets and dynamic implications. *Nature*, **461**, 1227–1233.

Alexander, C. M. O'D., Boss A. P. and Carlson R. W. (2001). The early evolution of the inner solar system: A meteoritic perspective. *Science* **293**, 64–68.

Alexander, C. M. O'D., Boss, A. P., Keller, L. P., Nuth, J. A. and Weinberger, A. (2007). Astronomical and meteoritic evidence for thermal processing of interstellar dust in protoplanetary disks. In *Protostars and Planets V*, ed. B. Reipurth, D. Jewitt and K. Keil. Tucson, AZ: University of Arizona Press, pp. 801–813.

Armitage, P. J. (2011). Dynamics of protoplanetary disks. *Annu. Rev. Astron. Astrophys.*, **49**, 195–236.

Asphaug, E. (2010). Similar-sized collisions and the diversity of planets. *Chemie der Erde-Geochemistry* **70**, 199–219.

Asphaug, E. (2014). Impact origin of the Moon? *Annu. Rev. Earth Planet. Sci.,* **42**, 551–578.

Asphaug, E. and Reufer, A. (2014). Mercury and other iron-rich planetary bodies as relics of inefficient accretion. *Nature Geosci.*, **7**, 564–568. doi:10.1038/ngeo2189

Asphaug, E., Agnor, C. B. and Williams, Q. (2006). Hit-and-run planetary collisions. *Nature,* **439**, 155–160.

Badro, J., Brodholt, J. P., Siebert, J. and Ryerson, F. J. (2015). Core formation and core composition from coupled geochemical and geophysical constraints. *Proc. Natl. Acad. Sci.*, **112**, 12310–12314.

Barclay, T., Rowe, J. F., Lissauer, J. J., Huber, D., Fressin, F., Howell, S. B., Bryson, S. T., Chaplin, W. J., Désert, J-M., Lopez, E. D., Marcy, G. W., Mullaly, F., Ragozzine, D., Torres, G., Adams, E. R., Agol, E., Barrado, D., Basu, S., Bedding, T. R., Buchhave, L. A., Charbonneau, D., Christiansen, J. L., Christensen-Dalsgaard, J., Ciardi, D., Cochran, W. D., Dupree, A. K., Elsworth, Y., Everett, M., Fischer, D. A., Ford, E. B., Fortney, J. J., Geary, J. C., Haas, M. R., Handberg, R., Hekker, S., Henze, C. E., Horch, E., Howard, A. W., Hunter, R. C., Isaacson, H., Jenkins, J. M., Karoff, C., Kawaler, S. D., Kjeldsen, H., Klaus, T. C., Latham, D. W., Li, J., Lillo-Box, J., Lund, M. N., Lundkvist, M., Metcalfe, T. S., Miglio, A.,

page 26 of 37Morris, R. L, Quintana, E. V., Stello, D., Smith, J. C., Still, M. and Thompson, S. E. (2013) A sub-Mercury-sized exoplanet. *Nature*, **494**, 452–454.

Barnes, R., Gozdziewski, K. and Raymond, S. N. (2008). The successful prediction of the extrasolar planet HD 74156d. *Astrophys. J.*, **680**, L57–L60.

Batalha, N. M., Borucki, W. J., Bryson, S. T., Buchhave, L. A., Caldwell, D. A., Christensen-Dalsgaard J., Ciardi, D., Dunham, E. W., Fressin, F., Gautier III, T. N., Gilliland, R. L., Haas, M. R., Howell, S. B., Jenkins, J. M., Kjeldsen, H., Koch, D. G., Latham, D. W., Lissauer, J. J., Marcy, G. W., Rowe, J. F., Sasselov, D. D., Seager, S., Steffen, J. H., Torres, G., Basri, G. S., Brown, T. M., David Charbonneau, D., Christiansen, J., Clarke, B., Cochran, W. D., Dupree, A., Fabrycky, D. C., Fischer, D., Ford, E. B., Fortney, J., Girouard, F. R., Holman, M. J., Johnson, J., Isaacson, H., Klaus, T. C., Machalek, P., Moorehead, A. V., Morehead, R. C., Ragozzine, D., Tenenbaum, P., Twicken, J., Quinn, S., VanCleve, J., Walkowicz, L. M., Welsh, W. F., Devore, E. and Gould, A. (2011). Kepler's first rocky planet: Kepler-10b. *Astrophys. J.*, **729**, 27–48. doi:10.1088/0004-637X/729/1/27

Batygin, K. and Brown, M. E. (2010). Early dynamical evolution of the solar system: Pinning down the initial conditions of the Nice model. *Astrophys. J.*, **716**, 1323–1331.

Beckett, J. R. (1986). The origin of calcium-, aluminum-rich inclusions from carbonaceous chondrites: An experimental study. Ph.D. thesis, Univ. Chicago, 373 pp.

Benz, W., Slattery, W. L. and Cameron, A. G. W. (1988). Collisional stripping of Mercury's mantle. *Icarus*, **74**, 516–528.

Benz, W., Anic, A., Horner, J. and Whitby, J. A. (2007). The origin of Mercury. *Space Sci. Rev.*, **132**, 189–202.

Bonsor, A., Leinhardt, Z. M., Carter, P. J., Elliott, T., Walter, M. J. and Stewart, S. T. (2015). A collisional origin to Earth's non-chondritic composition? *Icarus*, **247**, 291–300.

Borucki, W. J. (2016). KEPLER mission: Development and overview. *Reports on Progress in Physics*, **79**, 036901 (40 pp.).

Bouvier, A. and Boyet, M. (2016) Primitive Solar System materials and Earth share a common initial $^{142}$Nd abundance. *Nature,* **537**, 399–402.

Boyet, M. and Carlson, R. W. (2005). $^{142}$Nd evidence for early (>4.53 Ga) global differentiation of the silicate Earth. *Science*, **309**, 576–581.

Boynton, W. V., Sprague, A. L., Solomon, S. C., Starr, R. D., Evans, L. G., Feldman, W. C., Trombka, J. I. and Rhodes, E. A. (2007). MESSENGER and the chemistry of Mercury's surface. *Space Sci. Rev.*, **131**, 85–104.

Braden, S. E. and Robinson, M. S. (2013). Relative rates of optical maturation of regolith on Mercury and the Moon. *J. Geophys. Res. Planets*, **118**, 1903–1914. doi:10.1002/jgre.20143

Bradley, J. P. (2014). Early solar nebula grains - interplanetary dust particles. In *Meteorites and Cosmochemical Processes*, ed. A. M. Davis, *Treatise on Geochemistry,* 2nd ed., Vol. 1, ed. H. D. Holland and K. Turekian. Amsterdam, Netherlands: Elsevier, pp. 287–308.

Brett, R. and Sato, M. (1984). Intrinsic oxygen fugacity measurements on seven chondrites, a pallasite, and a tektite and the redox state of meteorite parent bodies. *Geochim. Cosmochim. Acta,* **48**, 111–120.

Brownlee, D. (2014). The Stardust mission: Analyzing samples from the edge of the solar system. *Annu. Rev. Earth Planet. Sci.*, **42**, 179–205.

Buchwald, V. F. (1975). *Handbook of Iron Meteorites, Their History, Distribution, Composition and Structure*. Berkeley, CA: Univ. California Press. 1426 pp.

Burbine, T. H., Meibom, A. and Binzel, R. P. (1996). Mantle material in the main belt: Battered to bits? *Meteorit. Planet. Sci.*, **31,** 607–620.

Burkhardt, C. (2014). Isotopic composition of the Moon and the lunar isotopic crisis. In *Encyclopedia of Lunar Science*. Springer (online) 10.1007/SpringerReference_440362.

Burkhardt, C., Kleine, T., Bourdon, B., Palme, H., Zipfel, J., Friedrich, J. and Ebel, D. S. (2008). Hf-W systematics of Ca-Al-rich inclusions from carbonaceous chondrites: Dating the age of the solar system and core formation in asteroids. *Geochim. Cosmochim. Acta*, **72**, 6177–6197.

Burkhardt C., Borg, L. E., Brennecka, G. A., Shollenberger, Q. R., Dauphas, N. and Kleine T. (2016). A nucleosynthetic origin for the Earth's anomalous $^{142}$Nd composition. *Nature*, **537**, 394–398.

Busemann, H., Nguyen, A. N., Cody, G. D., Hoppe, P., Kilcoyne, A. L. D., Stroud, R. M., Zega, T. J. and Nittler, L. R. (2009). Ultra-primitive interplanetary dust particles from the comet 26P/Grigg-Skjellerup dust stream collection. *Earth Planet. Sci. Lett.*, **288**, 44–57.

Cameron, A. G. W. (1985). The partial volatilization of Mercury. *Icarus*, **64**, 285–294.

Cameron, A. G. W., Fegley, B., Benz, W. and Slattery, W. L. (1988). The strange density of Mercury: Theoretical considerations. In *Mercury*, ed. F. Vilas, C. R. Chapman and M. S. Matthews. Tucson, AZ: University of Arizona Press, pp. 692–708.

Canup, R. M. (2004). Dynamics of lunar formation. *Annu. Rev. Astron. Astrophys.*, **42**, 441–475.

Canup, R. M. (2008). Accretion of the Earth. *Phil. Trans. Roy. Soc. London A,* **366**, 4061–4075.

Canup, R. M. (2012). Forming a Moon with an Earth-like composition via a giant impact. *Science*, **338**, 1052–1055.

Canup, R. M. and Asphaug, E. (2001). Origin of the Moon in a giant impact near the end of the Earth's formation. *Nature*, **412**, 708–712.

Canup, R. M., Visscher, C., Salmon, J. and Fegley, B., Jr. (2015). Lunar volatile depletion due to incomplete accretion within an impact-generated disk. *Nature Geosci.,* **8**, 918–921. doi:10.1038/ngeo2574

Carry, B. (2012). Density of asteroids. *Planet. Space Sci.*, **73**, 98–118.

Carter, P. J., Leinhardt, Z. M., Elliott, T., Walter, M. J. and Stewart, S. T. (2015). Compositional evolution during rocky protoplanet accretion. *Astrophys. J.,* **813**, 72–91.

Chambers, J. E. (2004). Planetary accretion in the inner Solar System. *Earth Planet. Sci. Lett.,* **223**, 241–252.

Chambers, J. E. (2009a). Planetary migration: What does it mean for planet formation? *Annu. Rev. Earth Planet. Sci.*, **37**, 321–344.

Chambers, J. E. (2009b). An analytical model for the evolution of a viscous, irradiated disk. *Astrophys. J.* **705**, 1206–1214.

Chambers, J. E. (2013). Late-stage planetary accretion including hit-and-run collisions and fragmentation. *Icarus,* **224**, 43–56.

Chambers, J. E. (2014). Giant planet formation with pebble accretion. *Icarus*, **233**, 83–100.

Chapman, C. R. (1988). Mercury: Introduction to an end-member planet. In *Mercury*, ed. F. Vilas, C. R. Chapman and M. S. Matthews. Tucson, AZ: University of Arizona Press, pp. 1–23.

Chen, J. and Kipping, D. (2016). Probabilistic forecasting of the masses and radii of other worlds. *Astrophys. J.*, **834**, 17–30.

Clayton, R. N. and Mayeda, T. K. (1996). Oxygen isotope studies of achondrites. *Geochim. Cosmochim. Acta*, **60**, 1999–2017.

Consolmagno, G. J. and Britt, D. T. (2004). Meteoritical evidence and constraints on asteroid impacts and disruption. *Planet. Space Sci.*, **52**, 1119–1128.

Cottrell, E. and Kelley, K. A. (2011). The oxidation state of Fe in MORB glasses and the oxygen fugacity of the upper mantle. *Earth Planet. Sci. Lett.*, **305**, 270–282.

Ćuk, M. and Stewart, S. T. (2012). Making the Moon from a fast-spinning Earth: A giant impact followed by resonant despinning. *Science*, **338,** 1047–1052. doi:10.1126/science.1225542

Cuzzi, J. N., Hogan, R. C. and Shariff, K. (2008). Toward planetesimals: Dense chondrule clumps in the protoplanetary nebula. *Astrophys. J.*, **687**, 1432–1447.

Dahl, T. W. and Stevenson, D. J. (2010). Turbulent mixing of metal and silicate during planet accretion – and interpretation of the Hf–W chronometer. *Earth Planet. Sci. Lett.,* **295**, 177–186.

D'Alessio, P., Calvet, N., Hartmann, L., Lizano, S. and Cantó, J. (1999). Accretion disks around young objects. II. Tests of well-mixed models with ISM dust. *Astrophys. J.*, **527**, 893–909.

Dauphas, N. and Morbidelli, A. (2014). Geochemical and planetary dynamical views on the origin of Earth's atmosphere and oceans. In *The Atmosphere – History*, ed. J. Farquhar, *Treatise on Geochemistry,* 2nd ed., Vol. 6, ed. H. D. Holland and K. Turekian. Amsterdam: Netherlands: Elsevier, pp. 1–35.

Davis, A. M. (2006). Volatile evolution and loss. In *Meteorites and the Early Solar System II*, ed. D. Lauretta and H. Y. McSween, Jr. Tucson, AZ: University of Arizona Press, pp. 295–307.

Domingue, D. L., Chapman, C. R., Killen, R. M., Zurbuchen, T. H., Gilbert, J. A., Sarantos, M., Benna, M., Slavin, J. A., Schriver, D., Trávníček, P. M., Orlando, T. M., Sprague, A. L., Blewett, D. T., Gillis-Davis, J. J., Feldman, W. C., Lawrence, D. J., Ho, G. C., Ebel, D. S., Nittler, L. R., Vilas, F., Pieters, C. M., Solomon, S. C., Johnson, C. L., Winslow, R. M., Helbert, J., Peplowski, P. N., Weider, S. Z., Mouawad, N., Izenberg, N. R. and McClintock, W. E. (2014). Mercury's weather-beaten surface: Understanding Mercury in the context of lunar and asteroidal space weathering studies. *Space Sci. Rev.*, **181**, 121–214. doi:10.1007/s11214-014-0039-5

Dressing, C. D. and Charbonneau, D. (2015). The occurrence of potentially habitable planets orbiting M dwarfs estimated from the full Kepler dataset and an empirical measurement of the detection sensitivity. *Astrophys. J.*, **807**, 45–68.

Dwyer, C. A., Nimmo, F. and Chambers, J. E. (2015). Bulk chemical and Hf–W isotopic consequences of incomplete accretion during planet formation. *Icarus,* **245**, 145–152.

Ebel, D. S. (2001). Vapor/liquid/solid equilibria when chondrites collide. *Meteorit. Planet. Sci. Suppl.*, **36**, A52–A53.

Ebel, D. S. (2006). Condensation of rocky material in astrophysical environments. In *Meteorites and the Early Solar System II*, ed. D. Lauretta and H. Y. McSween, Jr. Tucson, AZ: University of Arizona Press, pp. 253–277.

Ebel, D. S. and Alexander, C. M. O'D. (2011). Equilibrium condensation from chondritic porous IDP enriched vapor: Implications for Mercury and enstatite chondrite origins. *Planet. Space. Sci.*, **59**, 1888–1894.

Ebel, D. S. and Grossman, L. (2000). Condensation in dust-enriched systems. *Geochim. Cosmochim. Acta*, **64**, 339–366.

Ebel, D. S. and Sack, R. O. (2013). Djerfisherite: Nebular source of refractory potassium. *Contrib. Mineral. Petrol.,* **166**, 923–934.

Ebel, D. S., Brunner, C., Leftwich, K., Erb, I., Lu, M., Konrad, K., Rodriguez, H., Friedrich, J. M. and Weisberg, M. K. (2016). Abundance, composition and size of inclusions and matrix in CV and CO chondrites. *Geochim. Cosmochim. Acta*, **172**, 322–356. doi:10.1016/j.gca.2015.10.007

Elkins-Tanton, L. T. (2013). Planetary science: Occam's origin of the Moon. *Nature Geosci.*, **6**, 996–998.

Ernst, C. M., Murchie, S. L., Barnouin, O. S., Robinson, M. L., Denevi, B. W., Blewett, D. T., Head, J. W., Izenberg, N. R., Solomon, S. C. and Roberts, J. H. (2010). Exposure of spectrally distinct material by impact craters on Mercury: Implications for global stratigraphy. *Icarus*, **209**, 210–223. doi:10.1016/j.icarus.2010.05.022

Espaillat, D., Muzerolle, J., Najita, J., Andrews, S., Zhu, Z., Calvet, N., Kraus, S., Hashimoto, J., Kraus, A. and D'Alessio, P. (2014). An observational perspective of transitional disks. In *Protostars and Planets VI*, ed. H. Beuther, R. Klessen, C. Dullemond and Th. Henning. Tucson, AZ: University of Arizona Press, pp. 497–520.

Evans, L. G., Peplowski, P. N., Rhodes, E. A., Lawrence, D. J., McCoy T. J., Nittler L. R., Solomon S. C., Sprague A. L., Stockstill-Cahill, K. R., Starr R. D., Weider S. Z., Boynton, W. F., Hamara, D. K. and Goldsten, J. O. (2012). Major-element abundances on the surface of Mercury: Results from the MESSENGER Gamma-Ray Spectrometer. *J. Geophys. Res.*, **117**, E00L07. doi:10.1029/2012JE004178

Evans L. G., Peplowski P. N., McCubbin F. M., McCoy T. J., Nittler L. R., Zolotov M. Yu., Ebel D. S., Lawrence D. J., Starr R. D., Weider S. Z. and Solomon S. C. (2015). Chlorine on the surface of Mercury: MESSENGER gamma-ray measurements and implications for the planet's formation and evolution. *Icarus*, **257**, 417–427.

Fabrycky, D. C., Lissauer, J. J., Ragozzine, D., Fowe, J. F., Steffen, J. H., Agol, E., Barclay, T., Batalha, N., Borucki W. and Ciardi, D. R. (2014). Architecture of Kepler's multi-transiting systems. II. New investigations with twice as many candidates. *Astrophys. J.*, **790**, 146–157.

Fang, J. and Margot, J. L. (2012). Architecture of planetary systems based on Kepler data: Number of planets and coplanarity. *Astrophys. J.*, **761**, 92–105.

Fedkin, A. V., Grossman, L., Humayun, M., Simon, S. B. and Campbell, A. J. (2015). Condensates from vapor made by impacts between metal-, silicate-rich bodies: Comparison with metal and chondrule in CB chondrites. *Geochim. Cosmochim. Acta*, **164**, 236–261.

Fegley, B., Jr. and Cameron, A. G. W. (1987). A vaporization model for iron/silicate fractionation in the Mercury protoplanet. *Earth Planet. Sci. Lett.* **82**, 207–222.

Feigelson, E. D. (2010). X-ray insights into star and planet formation. *Proc. Natl. Acad. Sci.*, **107**, 7153–7157.

Friedrich, J. M., Weisberg, M. K., Ebel, D. S., Biltz, A. E., Corbett, B. M., Iotzov, I. V., Khan, W. S. and Wolman, M. D. (2015). Chondrule size and density in all meteorite groups: A compilation and evaluation of current knowledge. *Chemie der Erde*, **75**, 419-443. doi:10.1016/j.chemer.2014.08.003

Frost, D. J., Mann, U., Asahara, Y. and Rubie, D. C. (2008). The redox state of the mantle during and just after core formation. *Phil. Trans. Roy. Soc. London A*, **366**, 4315–4337.

Genda, H., Kokubo, E. and Ida, S. (2011). Merging criteria for giant impacts of protoplanets. *Astrophys. J.,* **744**, 137–144.

Ghosal, S., Sack, R. O., Ghiorso, M. S. and Lipschutz, M. E. (1998). Evidence for a reduced, Fe-depleted martian mantle source region of shergottites. *Contrib. Mineral. Petrol.,* **130**, 346–357.

page 30 of 37Gladman, B. and Coffey, J. (2009). Mercurian impact ejecta: Meteorites and mantle. *Meteorit. Planet. Sci.*, **44**, 285–291.

Goldschmidt, V. M. (1937). The principles of distribution of chemical elements in minerals and rocks. The seventh Hugo Müller Lecture, delivered before the Chemical Society on March 17th, 1937. *J. Chem. Soc.*, **1937**, 655–673.

Grossman, J. N. (1988). Formation of chondrules. In *Meteorites and the Early Solar System*, ed. J. F. Kerridge and M. S. Matthews. Tucson, AZ: University of Arizona Press, pp. 680–696.

Grossman, L. and Larimer, J. W. (1974). Early chemical history of the solar system. *Rev. Geophys. Space Phys.*, **12**, 71–101.

Haisch, K. E., Lada, E. A. and Lada, C. J. (2001). Circumstellar disks in the IC 348 cluster. *Astrophys. J.*, **121**, 2065–2074.

Halliday, A. N. (2013). The origins of volatiles in the terrestrial planets. *Geochim. Cosmochim. Acta*, **105**, 146–171.

Hansen, B. M. S. (2009). Formation of the terrestrial planets from a narrow annulus. *Astrophys. J.*, **703**, 1131–1140.

Hartmann, L. (2009). The star-jet-disk system and angular momentum transfer. In *Protostellar Jets in Context*, ed. K. R. Tsinganos and M. Stute. Berlin: Springer, pp. 23–32.

Hauck S. A., II, Margot, J.-L., Solomon, S. C., Phillips, R. J., Johnson, C. L., Lemoine, F. G., Mazarico, E., McCoy, T. J., Padovan, S., Peale, S. J., Perry, M. E., Smith, D. E. and Zuber, M. T. (2013). The curious case of Mercury's internal structure. *J. Geophys. Res. Planets*, **118**, 1204–1220.

Hewins, R. H. and Ulmer, G. C. (1984). Intrinsic oxygen fugacities of diogenites and mesosiderite clasts. *Geochim. Cosmochim. Acta,* **48**, 1555–1560.

Hubbard, A. (2014). Explaining Mercury's density through magnetic erosion. *Icarus*, **241**, 329–335.

Izenberg, N. R., Klima, R. L., Murchie, S. L., Blewett, D. T., Holsclaw, G. M., McClintock, W. E., Malaret, E., Mauceri, C., Vilas, F., Sprague, A. L., Helbert, J., Domingue, D. L., Head III, J. W. Goudge, T. A., Solomon, S. C., Hibbitts, C. A. and Dyar, M. D. (2014). The low-iron, reduced surface of Mercury as seen in spectral reflectance by MESSENGER. *Icarus*, **228**, 364–374.

Jarosewich, E. (1990). Chemical analyses of meteorites: A compilation of stony and iron meteorite analyses. *Meteoritics*, **25**, 323–337

Johansen A., Blum J., Tanaka H., Ormel C., Bizzarro M. and Rickman H. (2014). The multifaceted planetesimal formation process. In *Protostars and Planets VI*, ed. H. Beuther, R. Klessen, C. Dullemond and Th. Henning. Tucson, AZ: University of Arizona Press, pp. 547–570.

Kant, I. (1755). *Universal Natural History and Theory of the Heavens*. In *Kant's Critical Religion* (2000), translated S. Palmquist. Aldershot: Ashgate, 320 pp.

Kargel, J. S. and Lewis, J. S. (1993). The composition and early evolution of Earth. *Icarus*, **105**, 1–25.

Keil, K. (1968). Mineralogical and chemical relationships among enstatite chondrites. *J. Geophys. Res.*, **73**, 6945–6976.

Kleine, T., Mezger, K., Palme, H. and Münker, C. (2004). The W isotope evolution of the bulk silicate Earth: Constraints on the timing and mechanisms of core formation and accretion. *Earth Planet. Sci. Lett.,* **228**,109–123.

Kleine, T., Touboul, M., Bourdon, B., Nimmo, F., Mezger, K., Palme, H., Jacobsen, S. B., Yin, Q.-Z. and Halliday, A. N. (2009). Hf-W chronology of the accretion and early evolution of asteroids and terrestrial planets. *Geochim. Cosmochim. Acta*, **73**, 5150–5188.

Klima, R. L., Izenberg, N. R., Murchie, S., Meyer, H. M., Stockstill-Cahill, K. R., Blewett, D. T., D'Amore, M., Denevi, B. W., Ernst, C. M., Helbert, J., McCoy, T., Sprague, A. L., Vilas, F., Weider, S. Z. and Solomon, S. C. (2013). Constraining the ferrous iron content of silicate minerals in Mercury's crust. *Lunar Planet. Sci.*, **44**, abstract 1602.

Kokubo, E. and Genda, H. (2010). Formation of terrestrial planets from protoplanets under a realistic accretion condition. *Astrophys. J. Lett.,* **714**, L21–L25.

Kokubo, E. and Ida, S. (1996). On runaway growth of planetesimals. *Icarus*, **123**, 180–191.

Kokubo, E. and Ida, S. (1998). Oligarchic growth of protoplanets. *Icarus*, **131**, 171–178.

Krauss, O. and Wurm, G. (2005). Photophoresis and the pile-up of dust in young circumstellar disks. *Astrophys. J.*, **630**, 1088–1092.

Krot, A. N., Amelin, Y., Cassen, P. and Meibom, A. (2005). Young chondrules in CB chondrites from a giant impact in the early Solar System. *Nature*, **436**, 989–992.

Krot, A. N., Amelin, Y., Bland, P., Ciesla, F. J., Connelly, H. J. Jr., Davis, A. M., Huss, G. R., Hutcheon, I. D., Makide, K., Nagashima, K., Nyquist, L. E., Russell, S. S., Scott, E. R. D., Thrane, K., Yurimoto, H. and Yin, Q.-Z. (2009). Origin and chronology of chondritic components: A review. *Geochim. Cosmochim. Acta*, **73**, 4963–4997.

Lambrechts, M. and Johansen, A. (2012). Rapid growth of gas-giant cores by pebble accretion. *Astron. Astrophys.*, 544, A32–A45.

Larimer, J. W. and Anders, E. (1967). Chemical fractionation in meteorites: II. Abundance patterns and their interpretation. *Geochim. Cosmochim. Acta*, **31**, 1239–1270.

Leinhardt, Z. M. and Stewart, S. T. (2011). Collisions between gravity-dominated bodies. I. Outcome regimes and scaling laws. *Astrophys. J.,* **745**, 79–106.

Leinhardt, Z. M., Dobinson, J., Carter, P. J. and Lines, S. (2015). Numerically predicted indirect signatures of terrestrial planet formation. *Astrophys. J.*, **806,** 23–32.

Levison, H. F., Morbidelli, A., Gomes, R. and Backman, D. (2007). Planet migration in planetesimal disks. In *Protostars and Planets V*, ed. B. Reipurth, D. Jewitt and K. Keil. Tucson, AZ: University of Arizona Press, pp. 669–684.

Levison, H. F., Kretke, K. A. and Duncan, M. J. (2015a). Growing the gas-giant planets by the gradual accumulation of pebbles. *Nature*, **524**, 322–324.

Levison, H. F., Kretke, K. A., Walsh, K. J. and Bottke W. F. (2015b). Growing the terrrestrial planets from the gradual accumulation of submeter-sized objects. *Proc. Natl. Acad. Sci.*, *112*, 14180–14185.

Lewis, J. S. (1972). Metal/silicate fractionation in the solar system. *Earth Planet. Sci. Lett.,* **15**, 286–290.

Lewis, J. S. (1973). Chemistry of the planets. *Annu. Rev. Phys. Chem.,* **24**, 339–352.

Lewis, J. S. (1988). Origin and composition of Mercury. In *Mercury*, ed. F. Vilas, C. R. Chapman and M. S. Matthews. Tucson, AZ: University of Arizona Press, pp. 651–666.

Lissauer, J. J., Fabrycky, D. C., Ford, E. B., Borucki, W. J., Fressin, F., Marcy, G. W., Rorosz, J. A., Rowe, J. F., Torres., G., Welsh, W. F., Batalha, N. M., Bryson, S. T., Buchhave, L. A., Caldwell, D. A., Cartre, J. A., Charbonneau, D., Christiansen, J. L., Cochran, W. D., Desert, J-M., Dunham, E. W., Fanelli, M. N., Fortney, J. J., Gautier, T. N. III, Geary, J. C., Gilliland, R. L., Haas, M. R., Hall, J. R., Holman, M. J., Coch, D. G., Latham, D. W., Lopez, E., McCauliff, S., Miller, N., Morehead, R. C., Quintana, E. V., Ragozzine, D., Sasselov, D.,

Short, D. R. and Stefffen, J. H. (2011). A closely packed system of low-mass, low-density planets transiting Kepler-11. *Nature*, **470**, 53–58.

Lock S. J., Stewart, S. T., Petaev, M. I., Leinhardt, Z. M., Mace, M., Jacobsen, S. B. and Ćuk, M. (2016). A new model for lunar origin: Equilibration with Earth beyond the hot spin stability limit. *Lunar Planet. Sci.*, **47**, abstract 2881.

Lodders, K. (2003). Solar system abundances and condensation temperatures of the elements. *Astrophys. J.*, **591**, 1220–1247.

Lodders, K. and Fegley, B. Jr. (1997). An oxygen isotope model for the composition of Mars. *Icarus*, **126**, 373–394.

Lodders, K. and Fegley, B. Jr. (1998). *The Planetary Scientist's Companion*. New York: Oxford University Press.

Lodders, K., Palme, H. and Gail, H. P. (2009). Abundances of the elements in the solar system. In *Landolt-Börnstein, New Series*, Vol. VI/4B, ed. J. E. Trümper. Berlin, Heidelberg, New York: Springer-Verlag, pp. 560–630.

Loesche, C., Wurm, G., Kelling, T., Teiser, J. and Ebel, D. S. (2016). The motion of chondrules and other particles in a protoplanetary disk with temperature fluctuations. *Mon. Not. Roy. Astron. Soc.*, **463**, 4167–4174.

Macke, R. J. (2012). *Survey of Meteorite Physical Properties: Density, Porosity and Magnetic Susceptibility*. Ph.D. thesis, University of Central Florida.

Mahoney, T. J. (2014). *Mercury, A Compendium of the Astronomical Lexicon, Part A: Gazetteer and Atlas of Astronomy, Vol. I: The Terrestrial Planets*, Part 1. New York: Springer. doi:10.1007/978-1-4614-7951-2

Malavergne, V., Toplis, M. J., Berthet, S. and Jones J. (2010). Highly reducing conditions during core formation on Mercury: Implications for internal structure and the origin of a magnetic field. *Icarus*, **206**, 199–209. doi:10.1016/j.icarus.2009.09.001

Marcus, R. A., Stewart, S. T., Sasselov, D. and Hernquist, L. (2009). Collisional stripping and disruption of super-Earths. *Astrophys. J. Lett.*, **700**, L118–L122.

Marcus, R. A., Sasselov, D., Hernquist, L. and Stewart, S. T. (2010a). Minimum radii of super-Earths: Constraints from giant impacts. *Astrophys. J. Lett.*, **712**, L73–L76.

Marcus, R. A., Sasselov, D., Stewart, S. T. and Hernquist, L. (2010b). Water/icy super-Earths: Giant impacts and maximum water content. *Astrophys. J. Lett.,* **719**, L45–L49.

McClure, M. K., D'Alessio, P., Calvet, N., Espaillat, C., Hartmann, L., Sargent, B., Watson, D. M., Ingleby, L. and Hernández, J. (2013). Curved walls: Grain growth, settling, and composition patterns in T Tauri disk dust sublimation fronts. *Astrophys. J.*, **775**, 114–124.

McCubbin, F. M., Riner, M. A., Vander Kaaden, K. E. and Burkemper L. K. (2012). Is Mercury a volatile-rich planet? *Geophys. Res. Lett.*, **39**, L09202. doi:10.1029/2012GL051711

McDonough W. F. (2014). Compositional model for the Earth's core. In *The Mantle and Core,* ed. R. W. Carlson, *Treatise on Geochemistry*, 2nd ed., Vol. 3, ed. H. D. Holland and K. Turekian. Amsterdam, Netherlands: Elsevier, pp. 559–577.

McNally, C. P. and McClure, M. K. (2017). Photophoretic levitation and trapping of dust in the inner regions of protoplanetary disks. *Astrophys. J.*, **834**, 48–60.

McNally, C. P., Hubbard, A., Mac Low, M-M., Ebel, D. S. and D'Alessio, P. (2013). Mineral processing by short circuits in protoplanetary disks. *Astrophys. J.*, **767**, L2–L7.

Melosh, H. J. (2014), New approaches to the Moon's isotopic crisis. *Phil. Trans. Roy. Soc. London A,* **372,** 20130168.

Messenger, S., Keller, L. P., Stadermann, F. J., Walker, R. M. and Zinner, E. (2003). Samples of stars beyond the Solar System: Silicate grains in interplanetary dust. *Science*, **300**, 105–108.

Mitchell, D. and de Pater, I. (1994). Microwave imaging of Mercury's thermal emission at wavelengths from 0.3 to 20.5 cm. *Icarus*, **110**, 2–32.

Morbidelli, A. and Raymond, S. N. (2016). Challenges in planet formation. *J. Geophys. Res. Planets*, **121**, 1962–1980. doi:10.1002/2016JE005088

Morbidelli, A., Tsiganis, K., Crida, A., Levison, H. F. and Gomes, R. (2007). Dynamics of the giant planets of the Solar System in the gaseous protoplanetary disk and their relationship to the current orbital architecture. *Astron. J.*, **134**, 1790–1798.

Morbidelli, A., Lunine, J. I., O'Brien, D. P., Raymond, S. N. and Walsh, K.J. (2012). Building terrestrial planets. *Annu. Rev. Earth Planet. Sci.*, **40**, 251–275.

Morgan, J. W. and Anders, E. (1980). Chemical composition of Earth, Venus, and Mercury. *Proc. Natl. Acad. Sci.*, **77**, 6973–6977.

Moriarty, J., Madhusudhan, N. and Fischer, D. (2014) Chemistry in an evolving protoplanetary disk: Effects on terrestrial planet composition. *Astrophys. J.*, **787**, 81–91.

Morishima, R., Golabek G. J. and Samuel, H. (2013). N-body simulations of oligarchic growth of Mars: Implications for Hf–W chronology. *Earth Planet. Sci. Lett.*, **366,** 6–16.

Murchie, S. L., Klima, R. L., Denevi, B. W., Ernst, C. M., Keller, M. R., Domingue, D. L., Blewett, D. T., Chabot, N. L, Hash, C. D., Malaret, E., Izenberg, N. R., Vilas, F., Nittler, L. R., Gillis-Davis-J. J., Head, J. W. and Solomon, S. C. (2015). Orbital multispectral mapping of Mercury with the MESSENGER Mercury Dual Imaging System: Evidence for the origins of plains units and low-reflectance material. *Icarus*, **254**, 287–305.

Nakajima, M. and Stevenson, D. J. (2015). Melting and mixing states of the Earth's mantle after the Moon-forming impact. *Earth Planet. Sci. Lett.*, **427**, 286–295.

Namur, O., Charlier, B., Holtz, F., Cartier, C. and McCammon, C. (2016). Sulfur solubility in reduced mafic silicate melts: Implications for the speciation and distribution of sulfur on Mercury. *Earth Planet. Sci. Lett.,* **448**, 102–114.

Nittler, L. R., Starr, R. D., Weider, S. Z., McCoy, T. J., Boynton, W. V., Ebel, D. S., Ernst, C. M., Evans, L. G., Goldsten, J. O., Hamara, D. K., Lawrence, D. J., McNutt, R. L. Jr., Schlemm, C. E. II, Solomon, S. C. and Sprague, A. L. (2011). The major-element composition of Mercury's surface from MESSENGER X-ray spectrometry. *Science*, **333**, 1847–1850.

O'Brien, D. P., Morbidelli A. and Levison, H. F. (2006). Terrestrial planet formation with strong dynamical friction. *Icarus,* **184**, 39–58.

O'Neill, H. St. C. and Palme H. (2008). Collisional erosion and the non-chondritic composition of the terrestrial planets. *Phil. Trans. Roy. Soc. London A*, **366**, 4205–4238.

Ostro, S. J., Hudson, R. S., Nolan, M. C., Margo, J-L., Scheeres, D. J., Campbell, D. B., Magri, C., Giorgini, J. D. and Yeomans, D. K. (2000). Radar observations of asteroid 216 Kleopatra. *Science*, **288**, 836–839. doi:10.1126/science.288.5467.836

Pasek, M. A., Milsom, J. A., Ciesla, F. J., Lauretta, D. S., Sharp C. M. and Lunine, J. I. (2005). Sulfur chemistry with time-varying oxygen abundance during Solar System formation. *Icarus*, **175**, 1–14.

Peplowski, P. N., Evans, L. G., Hauck, S. A. II, McCoy, T. J., Boynton, W. V., Gillis-Davis, J., Ebel, D. S., Goldsten, J. O., Hamara, D. K., Lawrence D. J., McNutt, R. L. Jr., Nittler, L. R., Solomon, S. C., Rhodes, E. A., Sprague, A. L., Starr, R. D. and Stockstill-Cahill, K. R. (2011). Radioactive elements on Mercury's surface from MESSENGER: Implications for the planet's formation and evolution. *Science*, **333**, 1850–1852.

Peplowski, P. N., Lawrence, D. J., Rhodes, E. A., Sprague, A. L., McCoy, T. J., Denevi, B. W., Evans, L. G., Head, J. W., Nittler, L. R., Solomon, S. C., Stockstill-Cahill, K. R. and Weider, S. Z. (2012). Variations in the abundances of potassium and thorium on the surface of Mercury: Results from the MESSENGER Gamma-Ray Spectrometer. *J. Geophys. Res.* **117**, E00L04. doi:10.1029/2012JE004141

Peplowski, P. N., Evans, L. G., Stockstill-Cahill, K. R., Lawrence, D. J., Goldsten, J. O., McCoy, T. J., Nittler, L. R., Solomon S. C., Sprague, A. L., Starr, R. D. and Weider, S. Z. (2014). Enhanced sodium abundance in Mercury's north polar region revealed by the MESSENGER Gamma-Ray Spectrometer. *Icarus,* **228**, 86–95.

Peplowski, P. N., Lawrence, D. J., Evans, L. G., Klima, R. L., Blewett, D. T., Goldsten, J. O., Murchie, S. L., McCoy, T. J., Nittler, L. R., Solomon, S. C., Starr, R. D. and Weider, S. Z. (2015). Constraints on the abundance of carbon in near-surface materials on Mercury: Results from the MESSENGER gamma-ray spectrometer. *Planet. Space Sci.*, **108**, 98–107.

Peplowski, P. N., Klima, R. L., Lawrence, D. J., Ernst, C. M., Denevi, B. W., Frank, E. A., Goldsten, J. O., Murchie, S. L., Nittler, L. R. and Solomon, S. C. (2016). Remote sensing evidence for an ancient carbon-bearing crust on Mercury, *Nature Geosci.*, **9**, 273–276. doi:10.1038/ngeo2669

Pignatale, F. C., Liffman, K., Maddison, S. T. and Brooks, G. (2016). 2D condensation model for the inner Solar Nebula: An enstatite-rich environment. *Mon. Not. Roy. Astron. Soc.*, **457**, 1359–1370.

Quintana, E. V., Barclay, T., Borucki, W., Rowe, J. F. and Chambers, J. E. (2016). Giant impacts on Earth-like worlds. *Astrophys. J.*, **821**, 126–139.

Raymond, S. N., O'Brien, D. P., Morbidelli, A. and Kaib, N. A. (2009). Building the terrestrial planets: Constrained accretion in the inner Solar System. *Icarus*, **203**, 644–662.

Reipurth, B. and Bally, J. (2001). Herbig-Haro flows: Probes of early stellar evolution. *Annu. Rev. Astron. Astrophys.*, **49**, 195–236.

Reufer, A., Meier, M. M. M., Benz, W. and Weiler, R. (2012). A hit-and-run giant impact scenario. *Icarus*, **221**, 296–299.

Righter, K. (2003). Metal-silicate partitioning of siderophile elements and core formation in the early Earth. *Annu. Rev. Earth Planet. Sci.*, **31**, 135–174.

Righter, K., Arculus, R. J., Delano, J. W. and Paslick C. (1990). Electrochemical measurements and thermodynamic calculations of redox equilibria in pallasite meteorites: Implications for the eucrite parent body. *Geochim. Cosmochim. Acta*, **54**, 1803–1815.

Righter, K., Drake, M. J. and Scott, E. (2006). Compositional relationships between meteorites and terrestrial planets. In *Meteorites and the Early Solar System II*, ed. D. Lauretta and H. Y. McSween, Jr. Tucson, AZ: University Arizona, pp. 803–828.

Righter, K., Humayun, M. and Danielson, L. (2008). Partitioning of palladium at high pressures and temperatures during core formation. *Nature Geosci.*, **1**, 321–323. doi:10.1038/ngeo180

Righter, K., Sutton, S. R., Danielson, L., Pando, K. and Newville, M. (2016). Redox variations in the inner solar system with new constraints from vanadium XANES in spinels. *Am. Mineral.*, **101**, 1928–1942.

Ringwood, A. E. and Kesson, S. E. (1977). Basaltic magmatism and the bulk composition of the Moon. *Moon*, **16**, 425–464.

Rizo, H., Walker, R. J., Carlson, R. W., Horan, M. F., Mukhopadhyay, S., Manthos, V., Francis, D. and Jackson, M. G. (2016). *Science*, **352**, 809–812.

Robinson, M. S. and Taylor, G. J. (2001). Ferrous oxide in Mercury's crust and mantle. *Meteorit. Planet. Sci.*, **36**, 841–847.

Robinson, M. S., Murchie, S. L., Blewett, D. T., Domingue, D. L., Hawkins, S. E., III, Head, J. W., Holsclaw, G. M., McClintock, W. E., McCoy, T. J., McNutt, R. L., Jr., Prockter, L. M., Solomon, S. C., Watters, T. R. (2008). Reflectance and color variations on Mercury: Regolith processes and compositional heterogeneity. *Science,* **321**, 66–69. doi:10.1126/science.1160080

Rubie, D. C., Jacobson, S. A., Morbidelli, A., O'Brien, D. P., Young E. D., de Vries, J., F. Nimmo, F., Palme, H. and Frost, D. J. (2015). Accretion and differentiation of the terrestrial planets with implications for the compositions of early-formed Solar System bodies and accretion of water. *Icarus*, **248**, 89–108.

Rudge, J. F., Kleine, T. and Bourdon, B. (2010). Broad bounds on Earth's accretion and core formation constrained by geochemical models. *Nature Geosci.,* **3**, 439–443.

Russell, S. S., Hartmann, L., Cuzzi J. N., Krot, A. N., Gounelle, M. and Weidenschilling S. J. (2006). Timescales of the solar protoplanetary disk. In *Meteorites and the Early Solar System II*, ed. D. Lauretta and H.Y. McSween, Jr. Tucson, AZ: University Arizona, pp. 233–251.

Safronov, V. S. (1972). *Evolution of the Protoplanetary Cloud and Formation of the Earth and Planets*. Tech. Transl. F-677. Washington, DC: NASA.

Sarid, G., Stewart, S. T. and Leinhardt, Z. M. (2014). Mercury, the impactor. *Lunar Planet. Sci.*, **46**, abstract 2723.

Schönbächler, M., Carlson, R. W., Horan M. F., Mock T. D. and Hauri, E. H. (2010). Heterogeneous accretion and the moderately volatile element budget of Earth. *Science*, **328**, 884–887.

Simon, S. B., Sutton, S. R. and Grossman, L. (2007). Valence of titanium and vanadium in pyroxene in refractory inclusion interiors and rims. *Geochim. Cosmochim. Acta*, **71**, 3098–3118.

Smith, J. V. (1979). Mineralogy of the planets: A voyage in space and time. *Mineral. Mag.*, **43**, 1–89.

Sneden, S., Lawler, J. E., Cowan, J. J., Ivans, I. I. and Den Hartog, E. A. (2009). New rare earth element abundance distributions for the sun and five r-process-rich very metal-poor stars. *Astrophys. J. Suppl. Ser*. **182**, 80–96.

Solomon, S. C., McNutt, R. L., Jr., Gold, R. E., Acuña, M. H., Baker, D. N., Boynton, W. V., Chapman, C. R., Cheng, A. F., Gloeckler, G., Head, J. W. III, Krimigis, S. M., McClintock, W. E., Murchie, S. L., Peale, S. J., Phillips, R. J., Robinson, M. S., Slavin, J. A., Smith, D. E., Strom, R. G., Trombka, J. I. and Zuber, M T. (2001). The MESSENGER mission to Mercury: Scientific objectives and implementation. *Planet. Space Sci.*, **49**, 1445–1465.

Solomon, S. C., McNutt, R. L., Jr., Gold, R. E. and Domingue, D. L. (2007). MESSENGER mission overview. *Space Sci. Rev.*, **131**, 3–39.

Sprague, A. L. and Roush, T. L. (1998). Comparison of laboratory emission spectra with Mercury telescopic data. *Icarus*, **133**, 174–183.

Stewart, S. T. and Leinhardt, Z. M. (2012). Collisions between gravity-dominated bodies. II. The diversity of impact outcomes during the end stage of planet formation. *Astrophys. J.*, **751**, 32–49.

Stewart, S. T., Leinhardt, Z. M. and Humayun, M. (2013). Giant impacts, volatile loss, and the K/Th ratios on the Moon, Earth, and Mercury. *Lunar Planet. Sci.*, **44**, abstract 2306.

Stewart, S. T., Lock, S. J., Petaev, M. I., Jacobsen, S. B., Sarid, G., Leinhardt, Z. M., Mukhopadhyay, S. and Humayun, M. (2016). Mercury impact origin hypothesis survives the volatile crisis: Implications for terrestrial planet formation. *Lunar Planet. Sci.,* **47**, abstract 2954.

Svetsov, V. (2011). Cratering erosion of planetary embryos. *Icarus*, 214, 316–326.

Szymanski, A., Brenker, F. E., Palme, H. and El Goresy, A. (2010). High oxidation state during formation of Martian nakhlites. *Meteorit. Planet. Sci.*, **45**, 21–31.

Tsiganis, K., Gomes, R., Morbidelli, A. and Levison, H. F. (2005). Origin of the orbital architecture of the giant planets of the Solar System. *Nature*, **435**, 459–461.

Tuff, J., Wade, J. and Wood, B. J. (2013). Volcanism on Mars controlled by early oxidation of the upper mantle. *Nature*, **498**, 342–345.

Urey, H. (1950). The origin and development of the Earth and other terrestrial planets. *Geochim. Cosmochim. Acta*, **1**, 209–277.

Vander Kaaden, K. E. and McCubbin, F. M. (2015). Exotic crust formation on Mercury: Consequences of a shallow, FeO-poor mantle. *J. Geophys. Res. Planets,* **120,** 195–209. doi:10.1002/2014je004733

Vilas, F. (1988). Surface composition of Mercury from reflectance spectrophotometry. In *Mercury*, ed. F. Vilas, C. R. Chapman and M. S. Matthews. Tucson, AZ: University of Arizona Press, pp. 59–76.

Vityazev, A. V., Pechernikova, G. V. and Safronov, V. S. (1988). Formation of Mercury and removal of its silicate shell. In *Mercury*, ed. F. Vilas, C. Chapman and M. S. Matthews. Tucson, AZ: University of Arizona Press, pp. 667-669.

Wadhwa, M. (2001). Redox state of Mars' upper mantle and crust from Eu anomalies in shergottite pyroxenes. *Science*, **291**, 1527–1530.

Wadhwa, M. (2008). Redox conditions on small bodies, the Moon and Mars. *Rev. Mineral. Geochem.* **68**, 493-510.

Wagner, F. W., Sohl, F., Hussmann, H., Grott, M. and Rauer, H. (2011). Interior structure models of solid exoplanets using material laws in the infinite pressure limit. *Icarus*, **214**, 366–376.

Walsh, K. J., Morbidelli, A., Raymond, S. N., O'Brein, D. P. and Mandell, A. M. (2011). A low mass for Mars from Jupiter's early gas-driven migration. *Nature*, **475**, 206–209. doi:10.1038/nature10201

Wänke, H. (1981) Constitution of terrestrial planets. *Phil. Trans. Roy. Soc. London A*, **303**, 287–302.

Warell, J. and Blewett, D. T. (2004). Properties of the Hermean regolith: V. New optical reflectance spectra, comparison with lunar anorthosites, and mineralogical modelling. *Icarus,* **168**, 257–276.

Wasson, J. T. and Kallemeyn, G. W. (1988). Composition of chondrites. *Phil. Trans. Roy. Soc. London A*, **325**, 535–544.

Wasson, J.T. and Wai, C. M. (1970). Composition of the metal, schreibersite and perryite of enstatite achondrites and the origin of enstatite chondrites and achondrites. *Geochim. Cosmochim. Acta*, **34**, 169–184.

Weidenschilling, S. J. (1978). Iron/silicate fractionation and the origin of Mercury. *Icarus*, **35**, 99–111.

Weider, S. Z., Nittler, L. R., Starr, R. D., McCoy, T. J., Stockstill-Cahill, K. R., Byrne, P. K., Denevi, B. W., Head, J. W. and Solomon, S. C. (2012). Chemical heterogeneity on Mercury's

page 37 of 37bibliographysurface revealed by the MESSENGER X-Ray Spectrometer. *J. Geophys. Res.,* **117**, E00L05. doi:10.1029/2012JE004153

Weider, S. Z., Nittler, L. R., Starr, R. D., McCoy, T. J. and Solomon, S. C. (2014). Variations in the abundance of iron on Mercury's surface from MESSENGER X-Ray Spectrometer observations. *Icarus,* **235**, 170–186.

Weider, S. Z., Nittler, L. R., Starr, R. D., Crapster-Pregont, E. J., Peplowski, P. N., Denevi, B. W., Head, J. W., Byrne, P. K., Hauck, S. A. II, Ebel, D. S. and Solomon, S. C. (2015). Evidence for geochemical terranes on Mercury: Global mapping of major elements with MESSENGER's X-Ray Spectrometer. *Earth Planet. Sci. Lett.*, **416**, 109–120.

Weisberg, M. K. and Kimura, M. (2012). The unequilibrated enstatite chondrites. *Chemie der Erde,* **72**, 101–115.

Weisberg, M. K., Prinz, M. and Nehru, C. E. (1988). Petrology of ALH85085: A chondrite with unique characteristics. *Earth Planet. Sci. Lett.*, **91**, 19–32.

Weisberg, M. K., Prinz, M., Clayton, R. N., Mayeda, T. K., Sugiura, N., Zashu, S. and Ebihara, M. (2001). A new metal-rich chondrite grouplet. *Meteorit. Planet. Sci.* **36**, 3401–3418.

Weisberg, M. K., Ebel, D. S., Nakashima, D., Kita, N. T. and Humayun, M. (2015). Petrology and geochemistry of chondrules and metal in NWA 5492 and GRO 95551: A new type of metal-rich chondrite. *Geochim. Cosmochim. Acta*, **167**, 269–285.

Weisberg M. K., Bigolski, J., Ebel, D. S. and Walker, D. (2016). Calcium-aluminum-rich (CAI) and sodium-aluminum-rich (NAI) inclusions in the PAT 91546 CH chondrite. *Lunar Planet. Sci.,* **47**, abstract 2152.

Wetherill, G. W. (1994). Provenance of the terrestrial planets. *Geochim. Cosmochim. Acta,* **58**, 4513–4520.

Williams, J. P. and Cieza, L. A. (2011). Protoplanetary disks and their evolution. *Annu. Rev. Astron. Astrophys.,* **49**, 67–118.

Wood, B. J., Walter, M. J. and Wade, J. (2006). Accretion of the Earth and segregation of its core. *Nature*, **441**, 825–833.

Wood, B. J., Wade, J. and Kilburn, M. (2009). Core formation and the oxidation state of the Earth: Additional constraints from Nb, V and Cr partitioning. *Geochim. Cosmochim. Acta*, **72**, 1415–1426.

Wurm, G., Trieloff, M. and Rauer, H. (2013). Photophoretic separation of metals and silicates: The formation of Mercury like planets and metal depletion in chondrites. *Astrophys. J.*, **769**, 78–85.

Yang J., Goldstein J. I. and Scott E. R. D. (2007). Iron meteorite evidence for early formation and catastrophic disruption of protoplanets. *Nature*, **446**, 888–891.

Zeng, L., Sasselov, D. and Jacobsen, S. (2016). Mass-radius relation for rocky planets based on PREM. *Astrophys. J.*, **819**, 127–131.

Zolotov, M. Yu., Sprague, A. L., Hauck, S. A., II, Nittler, L. R., Solomon, S. C. and Weider, S. Z. (2013). The redox state, FeO content, and origin of sulfur-rich magmas on Mercury. *J. Geophys. Res. Planets*, **118**, 138–146. doi:10.1029/2012JE004274